\documentclass[a4paper, twoside, twocolumn, 10pt]{book} 

\usepackage{graphicx,color}
\usepackage{multirow}
\usepackage{multicol} 
\usepackage{layout}
\usepackage{ulem} 
\usepackage{amsmath} 

\begin{document}


\begin{titlepage}
\Large \bf 
\begin{center}
Studies on the Structures and Physical Properties of \\
Crystal Polymorphs for Poly(Vinylidene Fluoride) \\
Based on the Density Functional Theory\\[2pc]
Ph.D Thesis\\
\vfil
\vfil
\vfil
Akira Itoh\\[2pc]\large {
\textit{Department of Chemical Sciences and Technology, \\Tokyo University of Science}\\[2pc]

\begin{tabular}{rl}
March 2014&(Japanese Version)\\
May 2014&(English Version)\\
\end{tabular}\\[4pc]}
\end{center}
\end{titlepage}

\tableofcontents 

\chapter{General Introduction}
\section{Introduction}

Poly(vinylidene fluoride) (PVDF) is attracting the attention of many researchers as the most popular ferroelectric polymer.

PVDF is a fluoropolymer developed in 1944, it has a high dielectric constant, high mechanical strength and high resistance to solvent. In 1969, Kawai discoverd strong piezoelectricity in its uniaxially-drawn and poled film. Soon afterward, pyroelectricity and second harmonic generation was found in the same polymer, PVDF attracted many researchers as a new functional polymer. In 1980, its ferroelectricity was proven by Furukawa \textit{et al.}, they observed $D$-$E$ hysteresis loops and high speed switching phenomena at lower temperature than its glass transition point applying high electric field.\cite{1Furu89} 

The theoretical prediction of the amount of spontaneous polarization of form I PVDF has been an important goal for many years. In table \ref{Ps studies}, important estimations are shown. 
The most primitive prediction canbe made simply by the summation of the "rigid" dipole moments of monomeric unit $\mu_r$ over a unit volume, $\mu_r$ can be estimate from the electric charge from electronegativity and structure of monomeric unit, summing up $\mu_r$ over a unit volume of the crystal yields
\begin{equation}
P_s = 2\mu_v / a b c = 130\,{\rm mC/m^2}
\end{equation}
where $a$, $b$ and $c$ are the lattice constants; $a=8.58\,{\rm \AA}$, $b=4.91\,{\rm \AA}$ and $c=2.56\,{\rm \AA}$.\cite{1Furu89, 1ThreeCrystal1}
In general, an electric dipole in crystal receives the local field, local field is the electric field made by surrounding dipoles, the real dipole moment must be influenced by the local field. 
In 1970 and 1975, Kakutani and Mopsik \textit{et al.} reported that $P_s$ with the Lorentz field theory yields 220$\,$mC/m$^2$, \cite{1Lorentz1, 1Lorentz2} however the theory of Lorentz field must be adopted in higher symmetry system than cubic system.
In 1982, Purvis and Taylor estimated $P_s = 86\,{\rm mC/m^2}$ with point dipole theory, \cite{1Purvis1} they assumed that a point dipole on the real base centered orthorhombic crystal is enlarge by local field estimated using Colpa's theory \cite{1Colpa1, 1Colpa2} via permittivity of the crystal.
In 1985, Al-Jishi and Taylor calculated $P_s = 127\,{\rm mC/m^2}$ with the model assuming that a pair of positive and negative charges separate a little distance on real Form I PVDF lattice, \cite{1Al-Jishi1} eventually this lead to a low local field and a $P_s$ close to that of the rigid dipole assumption.
Recently, more precise calculations with existing atoms and molecules are done.
In 1995, Carbeck \textit{et al.} computed molecular mechanics of Form I PVDF,\cite{1Carbeck1} this yield $P_s = 182\,{\rm mC/m^2}$.
In 2004, Nakhmans \textit{et al.} estimated $P_s = 178\,{\rm mC/m^2}$ with density functional theory (DFT), \cite{1Nakhmanson1} however this calculation neglect van der Waals interaction. \cite{1Nakhmanson2}

\begin{table*}\label{Ps studies} 
\caption{Studies of the amount of spontaneous polarization of Form I PVDF.}
\begin{center}
\begin{tabular}{rllrc}
\hline
&&
\multicolumn{1}{c}{Model}&
\multicolumn{1}{c}{$P_s$}&
\multicolumn{1}{c}{Ref.}\\
\hline 
&&Rigid dipoles&130 mC/m$^2$\\
1970&Kakutani&
\multirow{2}{*}{Lorentz field}&%
\multirow{2}{*}{220 mC/m$^2$}&\cite{1Lorentz1}\\
1975&Mopsik and Broadhurst&
&
&
\cite{1Lorentz2}\\
1982&Purvis and Taylor
&Point dipoles&86 mC/m$^2$&\cite{1Purvis1}\\
1985&Al-Jishi and Taylor&
Point charges&127 mC/m$^2$&\cite{1Al-Jishi1}\\
1995&Carbeck \textit{et al.}&
Molecular mechanics&182 mC/m$^2$&\cite{1Carbeck1}\\
2004&Nakhmanson \textit{et al.}&
Density functional theory&178 mC/m$^2$&\cite{1Nakhmanson1}\\
\hline
\end{tabular}
\end{center}
\end{table*}

\section{Poly(vinylidene fluoride)}
The functional properties of PVDF; piezoelectricity, pyroelectricity and ferroelectricity originate in the dipole moment of CH$_2$CF$_2$ unit in association with positive H and negative F.

One of the characteristic features of PVDF is the existence of at least four crystalline polymorphs, which are referred to as Forms I, \cite{1ThreeCrystal1} II,\cite{1ThreeCrystal1} III,\cite{gamma} IV \cite{delta} or $\beta$, $\alpha$, $\gamma$, $\delta$ phases, respectively. Form I consists of all-trans (TT) molecules packed in an antiparallel manner. Form II consists of trans-gauche-trans-gauche-minus ($\rm T G T \bar{G}$) molecules packed in an antiparallel manner. An intermediate conformation ($\rm T_3 G T_3 \bar{G}$) favors a parallel packing to generate Form III. There is a parallel version of Form II that is called Form IV. Regarding the molecular packing above stated, "parallel" and "antiparallel" are related to the dipole orientation in the direction perpendicular to the chain axis. Accordingly Forms I, III, IV are polar crystals and exhibit spontaneous polarization $P_s$. $\rm TGT\bar{G}$ and $\rm T_3 G T_3 \bar{G}$ conformations have a dipole component along the chain direction that generates an additional packing scheme, "up" and "down". It is believed that Forms II, III and IV adopt statistically random up-down packing. These crystalline polymorphs are illustrated in Figure

\begin{figure*}
\begin{center}
\begin{tabular}{ccc}
\multicolumn{1}{l}{(a)} &
\multicolumn{1}{l}{(b)} &
\multicolumn{1}{l}{(c)} \\
\\
\includegraphics[scale=0.3]{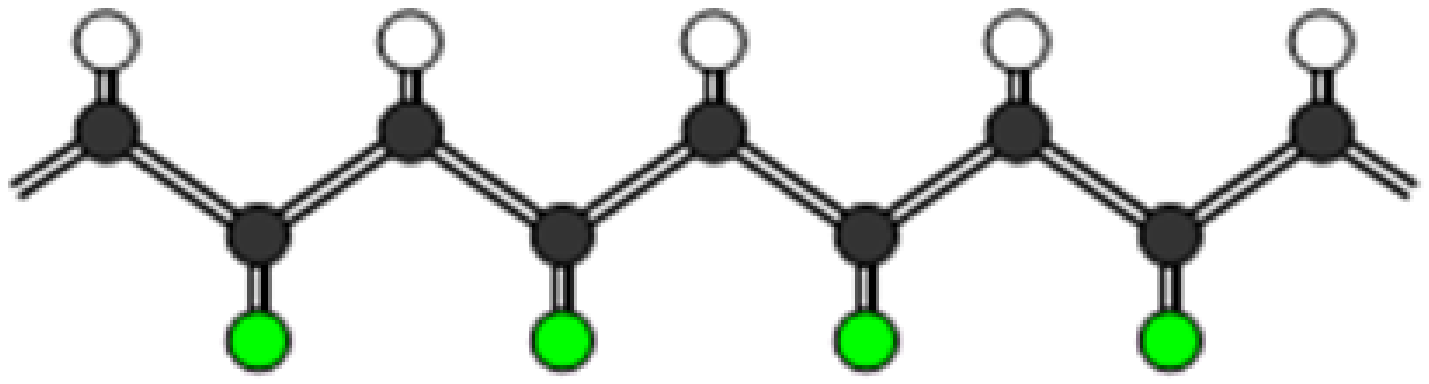}&
\includegraphics[scale=0.3]{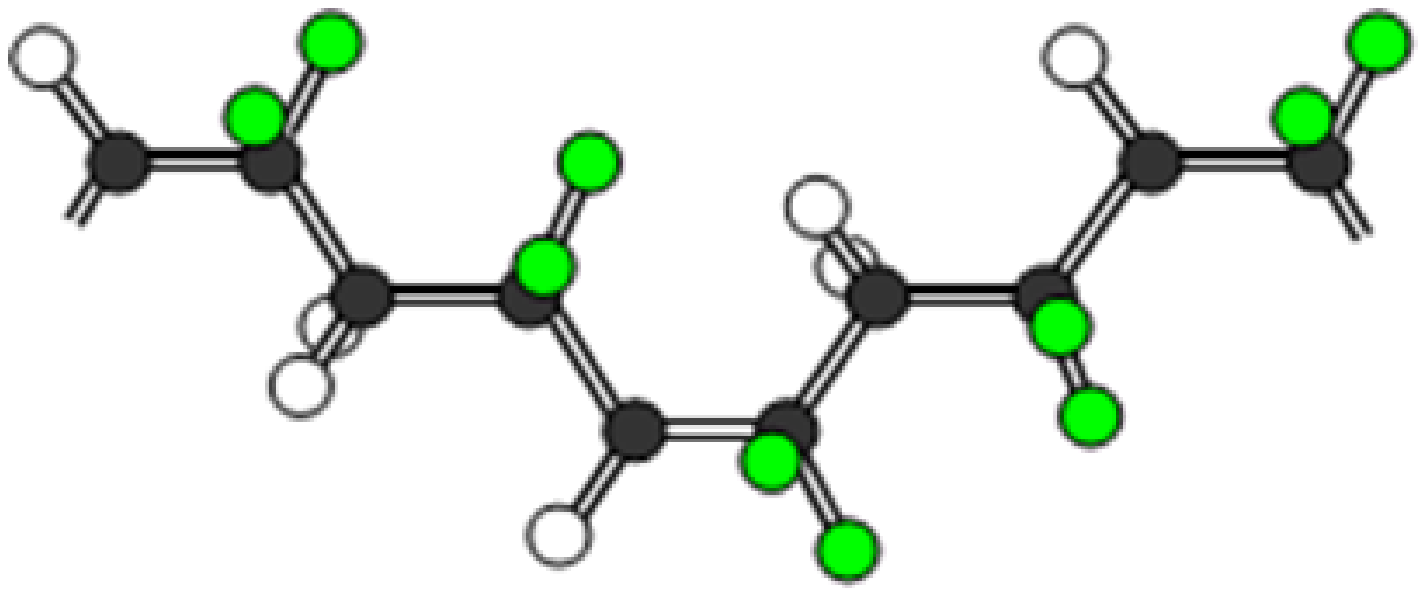}&
\includegraphics[scale=0.3]{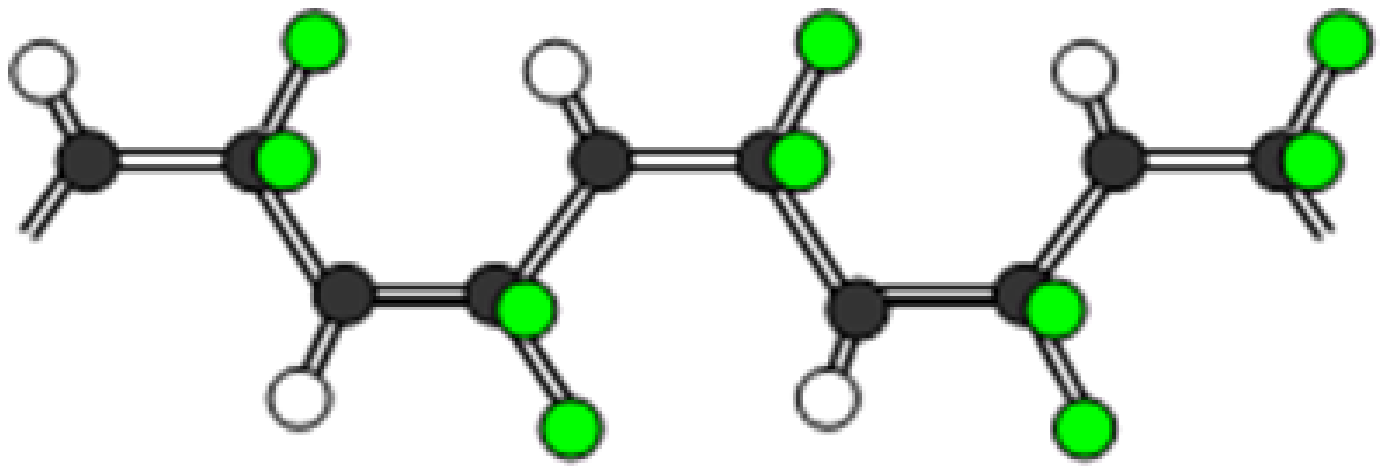}\\
All trans&$\rm T_3 GT_3 \bar{G}$&$\rm T GT \bar{G}$\\
\\
\\
\\
&&\multicolumn{1}{l}{(d)} \\
&&\includegraphics[scale=0.3]{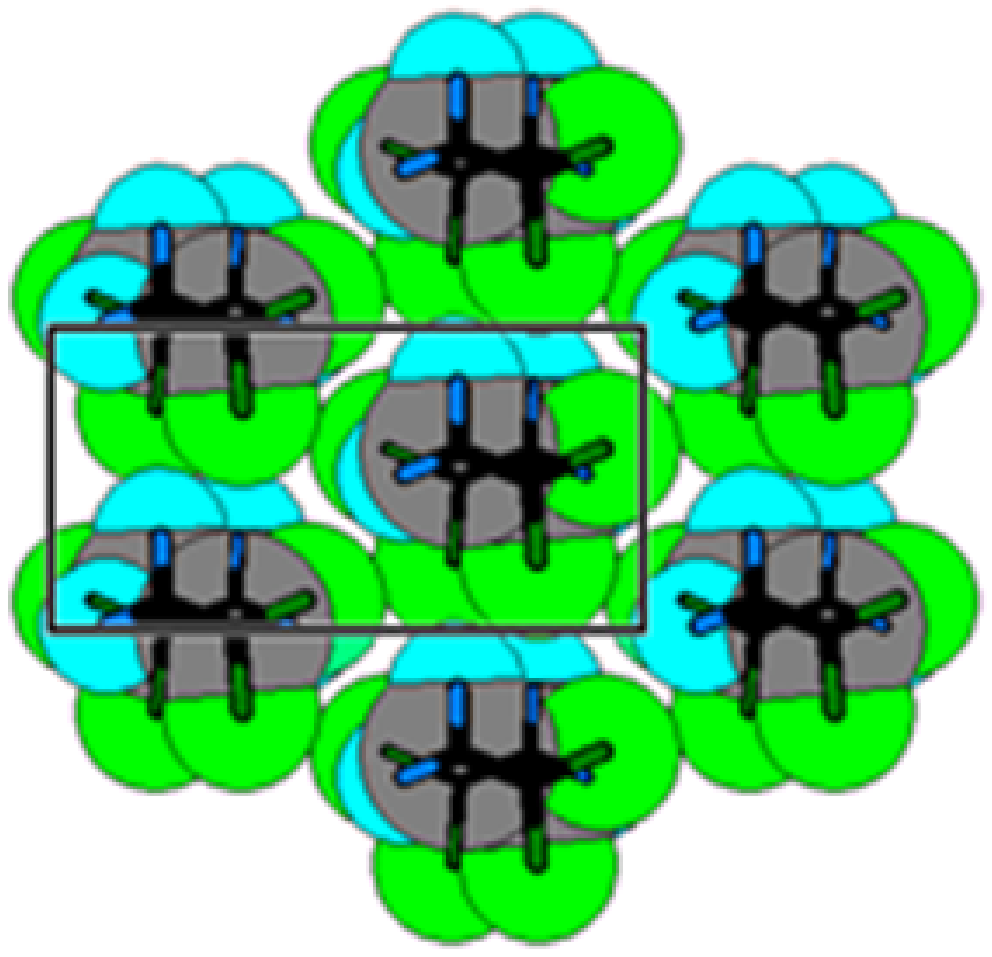}\\
&&Form II, $\rm TGT \bar{G}_{ad}$\\
\\
\\
\\
\multicolumn{1}{l}{(e)} &
\multicolumn{1}{l}{(f)}&
\multicolumn{1}{l}{(g)}\\
\includegraphics[scale=0.3]{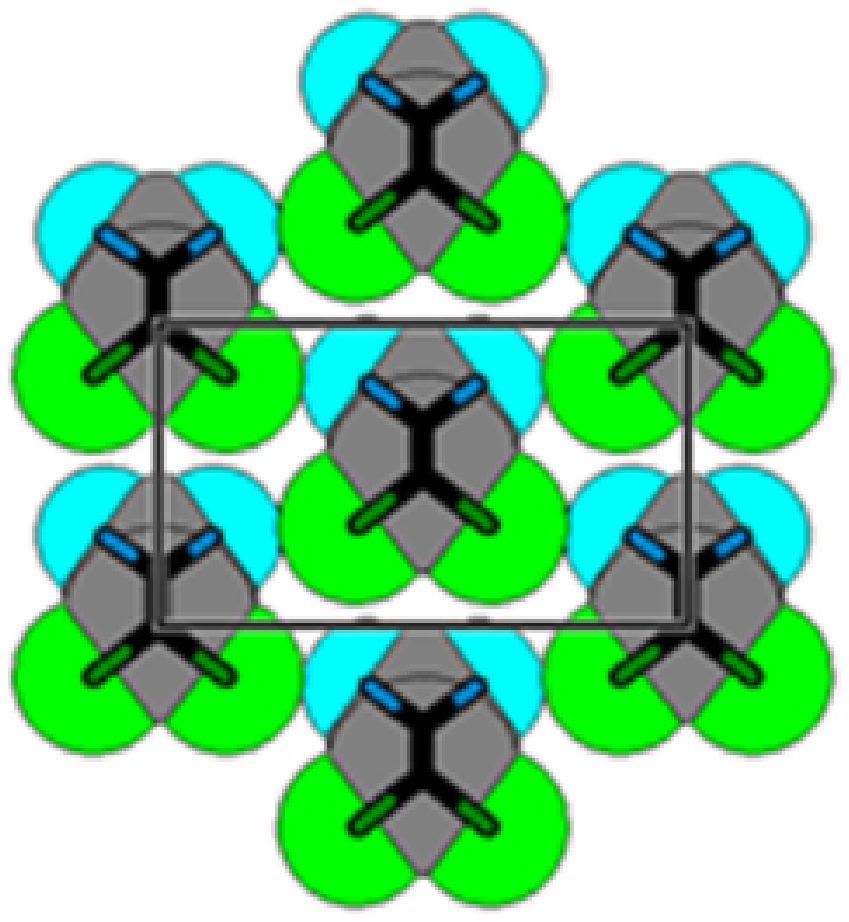}&
\includegraphics[scale=0.3]{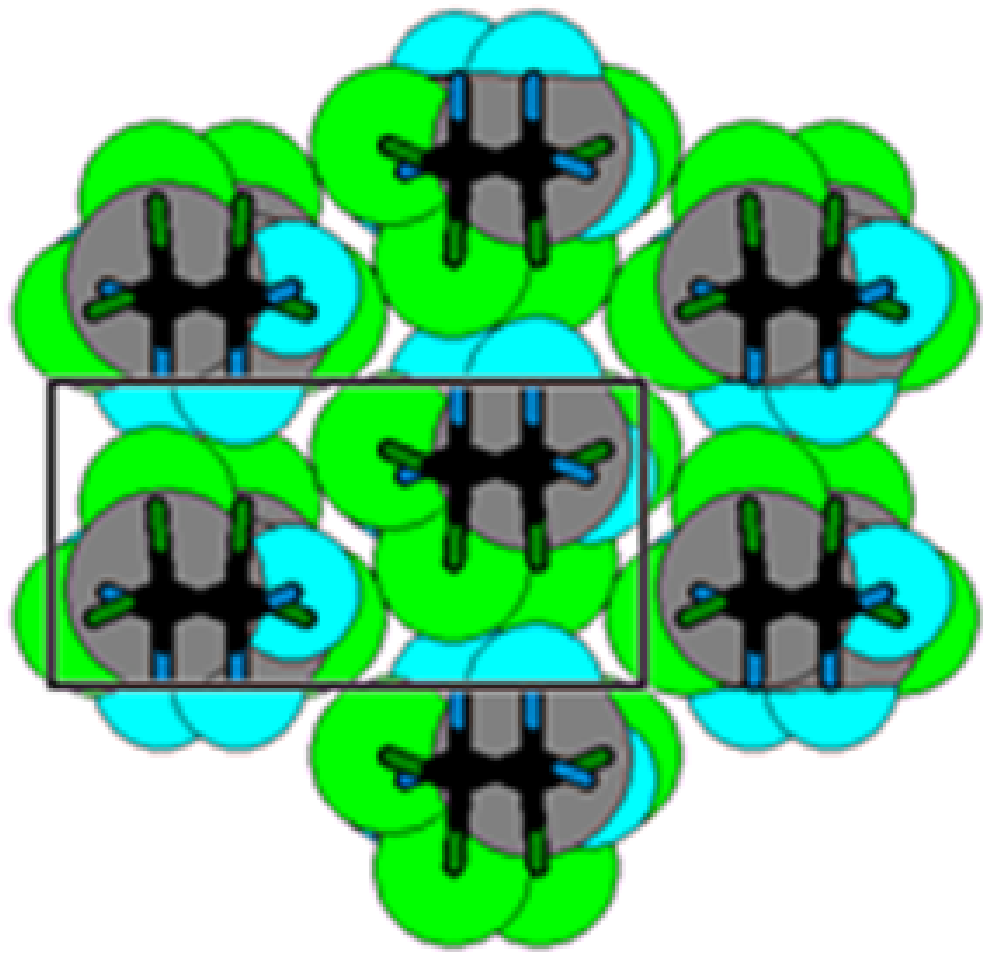}&
\includegraphics[scale=0.3]{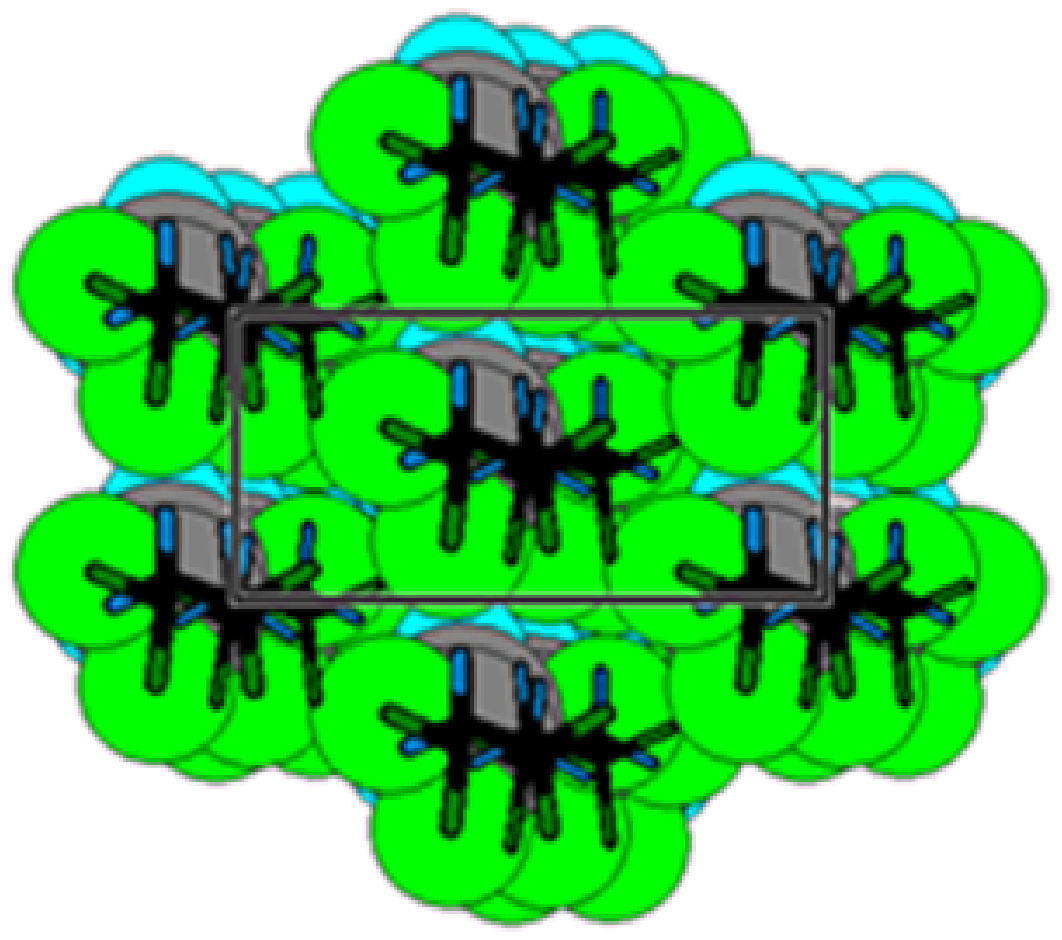}\\
Form I, $\rm TT_p$&
Form III, $\rm T_3 GT_3 \bar{G}_{pu}$&
Form IV, $\rm TGT \bar{G}_{pd}$\\
\end{tabular}
\end{center}
\caption{The three conformations of PVDF chains observed in different polymorphs. 
(a) TT, (b) $\rm T_3 GT_3 \bar{G}$, and (c) $\rm TGT\bar{G}$. 
The four PVDF crystalline polymorphs. 
Black rectangule in each crystal is the crystallographic unit cell. 
(d) Form II crystal of PVDF is non-polar one consisting from $\rm TGT\bar{G}$. 
(e) Form I cyrstal is polar TT. 
(f) Form III PVDF has polar $\rm T_3 GT_3 \bar{G}$ structure.
(g) Form IV is polar $\rm T GT \bar{G}$ crystal.
The suffix, "a" or "p" is packing style of vertical direction in figure, "a" means anti-parallel, "p" is parallel.
The other suffix "u" or "d" is placing style of chain direction, "up-up" and "up-down".
}\label{fig1}
\end{figure*}

These polymorphs are stable at room temperature but can be converted each other by applying external force, electric field and temperature changes. It is well known that melt-crystallization yields a nonpolar Form II and it can be converted into polar Form I by uniaxial-cold-drawing. The applied mechanical force causes a conformational change from alternately-twisted $\rm T G T \bar{G}$ in Form II into extended planar-zigzag TT which favors to be packed in a parallel manner to comprise Form II. Form I has the largest $P_s$ among three crystalline polymorphs of PVDF due to alignment of all molecular dipoles in one direction. Since the spontaneous polarization can be  switchable by action of a high electric field, Form I PVDF is ferroelectric. Heat treatment of Form II film under a certain temperature condition causes a conversion into Form III. An application of a high electric field leads to a conversion from Form II to Form IV. Both Forms III and IV are polar because of parallel packing. However, the properties of these polymorphs have not been understood well because the samples obtained by these treatments are usually mixtures of various polymorphs as well as noncrystalline regions.

Spontaneous polarization is the most critical quantity that may reflect ferroelectric cooperative interactions. It can be experimentally determined by means of $D$-$E$ hysteresis measurements. The amount of switched polarization is called a remanent polarization $P_r$ that is be close to $P_s$ if single-crystalline sample is used. This is not the case for PVDF. Likely to most crystalline polymers, the coexistence of noncrystalline regions is inevitable. Conventional Form I PVDF prepared by melt-crystallized and cold-drawn exhibit a crystallinity of about 50\%. The reported values of $P_r$ are scattered in the 50--80$\,\rm mC/m^2$ range. Special treatments such as high pressure crystallization and ultra-drawing can cause considerable increases in crystallinity and $P_r$ as well. The largest value of $P_r$ ever reported is ca. 100$\, \rm mC/m^2$.\cite{1Kana01} 

\section{Ferroelectrics}
A ferroelectric has spontaneous polarization which can be made reverse applying an external electric field. It is said that a material has a permanent dipole which can retain to one direction by intermolecular interaction or Coulomb interaction and can reverse applying an external field.

An electric dipole produces an electric field in the neighborhood and has an influence via Coulomb interactions each other. 
In a crystal, it is thought that a complicated local field is produced by the electric fields where it occurred by many electric dipoles.
According to the basic theory of ferroelectrics, spontaneous polarization is made by the ferroelectric interaction that an electric dipole is stabilizing each other through Coulomb interactions. 
In general ferroelectrics, dipoles influence each other via these process, this effect is called ferroelectric interaction. 
In this way, the spontaneous polarization reflects the interaction between electric dipoles and is the most important and essential quantity in ferroelectrics. 
In the real crystal, the non-localized electrons distribute complexly, the precise distribution must be evaluated to estimate local fields. 
In the case of PVDF, there is a possibility that the crystal structure of the molecular chain packing is merely retained by van der Waals interactions, there is discussions about the existence of the ferroelectric interactions.

PVDF performs the polarization reversal when a high electric field is applied. 
Form I PVDF has the highest remnant polarization (\textit{ca.} 100$\rm \,mC/m^2$) among the ferroelectric polymers. Against this experimental amount of the remnant polarization, the amount of the spontaneous polarization as the characteristic polarization of a certain ferroelectric crystal is the most important principal physical quantity. 

\begin{figure*}
\includegraphics[width=\textwidth]{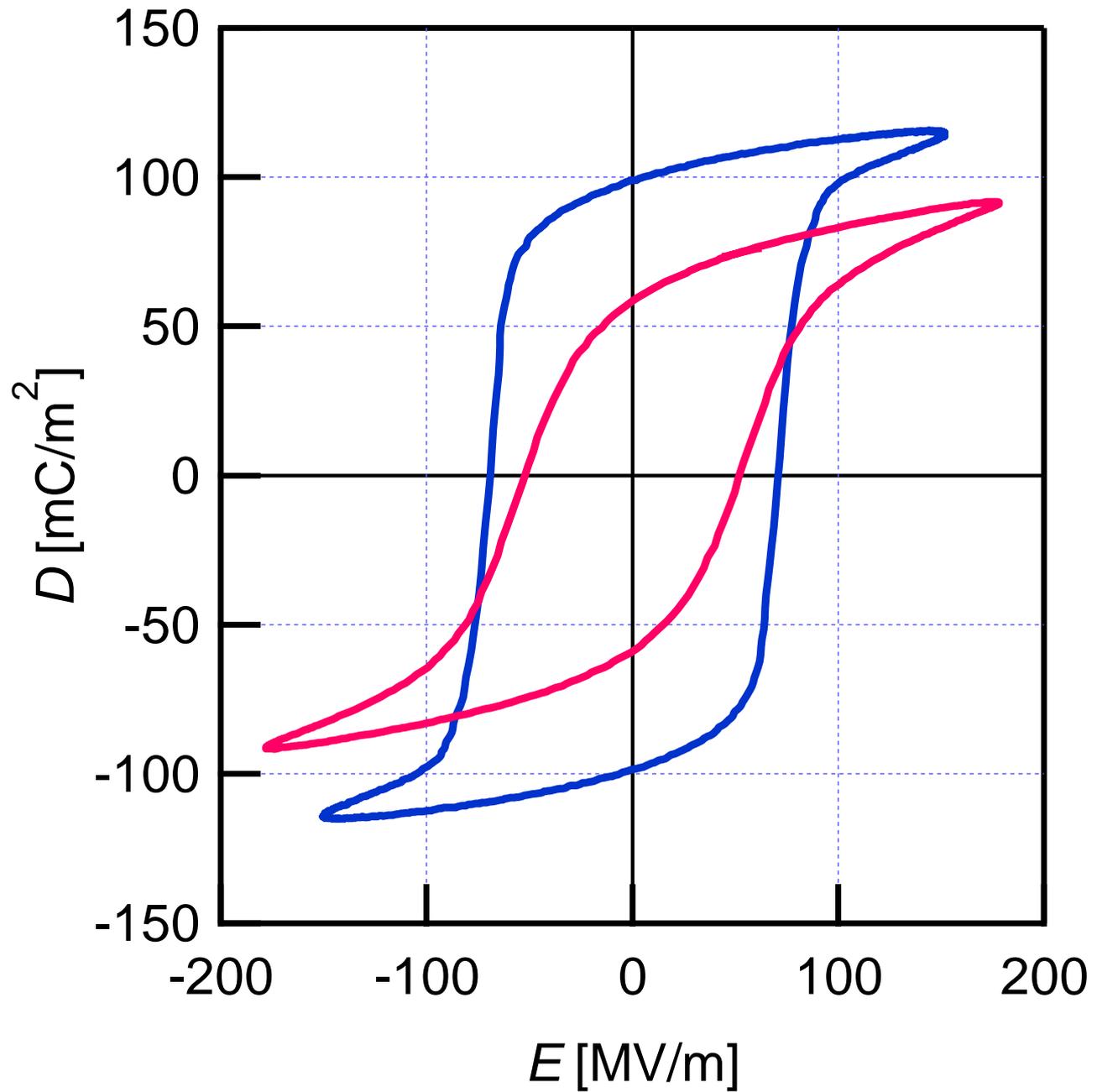}
\caption{Hysteresis loops of the electric displacement ($D$) versus electric field ($E$) of PVDF. The red line is the one of the typical sample (crystallinity \textit{ca.} 50\%). The blue line is one of co-extrusion drawing sample (crystallinity \textit{ca.} 80\%). \cite{1Kana01}}\label{hysteresis} 
\end{figure*}

A $D$-$E$ hysteresis loop is the chart of the electric displacement $D$ which indicates the polarization versus the applying electric field $E$. (Table \ref{hysteresis}) 
When the applying field is 0, the polarization is called the remnant polarization. 
When the polarization reversal is occurred, the electric field is named the coercive field. 
On the other hand, a ferroelectric has the phenomenon that it loses spontaneous polarization above Curie-Wise temperature. In the case of PVDF, the Curie-Wise point is above the melting point; therefore it can not be observed.

Ferroelectrics can be classified in two categories, displacement and order-disorder type. 
PVDF is a typical order-disorder ferroelectric. 
Also, the ferroelectricity of polymers was observed with odd-nylons and certain kind of polyamides. 
Especially in nylon-11, the thermal treatment dependency of its switching characteristic has been studied. 
Polarization reversal of nylon-11 occurs through the process that molecular chains are rotated after the intermolecular hydrogen bonds are cut temporarily. 
Therefore, it was reported that the narrower packing in crystal becomes, the more difficult polarization reversal will become. 
The displacement ferroelectrics, the other type of ferroelectrics, include ferroelectric ceramics consisting of the perovskite structure.

The properties of ferroelectrics can be described in a theory of phenomenology which treats polarization as an order coefficient. 
Some materials are called the indirect ferroelectrics, are distinguished from genuine ferroelectrics and have order coefficients except the polarization;
The relationship between the coefficient and the polarization causes ferroelectricity.
%
For example, DOBAMBC is a famous ferroelectric liquid-crystal. 
Most indirect ferroelectrics have permanent dipoles and have optical activity according to their chiral carbons. 
A molecule including a chiral carbon do not have mirror plains, therefore the spontaneous polarization is evolved by inducing a change of the polar structure with applying a mechanical strain.


%

If the amount of spontaneous polarization of PVDF was observed precisely and the components of dipole moment could be guessed, how the influence of ferroelectric interaction is working can be estimated. However, in the case of PVDF, a sample made from the each pure crystalline phase was not developed. 
About PVDF, application to nonvolatile memory is expected from the behavior of ferroelectrics that if a higher electric field than the coercive field was applied once, the polarization is sustained even if the field is 0.

\section{Purpose of This Study}
The spontaneous polarization of ferroelectrics is the essential quantity which reflects co-operative effect of electric dipoles, to evaluate it correctly, the local field which made by dipoles in the neighborhood must be treated. 
In the crystal, non-localized electrons distributes complexly. 
To derive the electronic distribution accurately is necessary to evaluate the local field correctly. 
In other words, valid electronic distribution must be derived with a quantum chemical calculation.

In this study, firstly the careful quantum chemical calculation will be performed to reproduce the lattice constants of the real crystals of PVDF to develop the precise computational method. 
Then, the suitability of the method will be evaluated through the process that some optical properties will be calculated, and the results will verify with the experimental values.
Finally, calculation of the amount of the spontaneous polarization of each crystal will be  performed and is the main aim.

About Form I PVDF, a number of precedent studies which include some quantum mechanical calculation reported about the amount of spontaneous polarization. 
Therefore, in this study, more careful calculations will be performed and will verify their calculations. 
In addition, the calculation about the ferroelectricity of Forms III and IV PVDF was not reported. 
For the first time, this study reports the amount of spontaneous polarization of these crystals. 

\section{Contents of This Thesis}
This thesis consists of six chapters.

Chapter 1 is "General Introduction", the summary of past studies and purpose of this study is described.

Chapter 2 is "Theoretical Backgrounds", the quantum chemical methods is employed  through the study especially the density functional theory and are described.

In Chapter 3 "Calculation of Form I PVDF", about the form I PVDF, using the agreements between computational result and a set of experimental values of the lattice constants, 
the each combination of various Hamiltonians and  and basis sets is tested. 
The Hamiltonians which are employed are  are various density functional methods and Hartree-Fock method. 

In chapter 4, "Crystalline and Electronic Structure of Polymorphs of PVDF". The author applied the calculation using the DFT method PBE0/cc-pVTZ to the all polymorphs of PVDF. The lattice constants of four crystalline polymorphs of PVDF were well reproduced, and the structure in the crystals is clarified. Packing energies, band structures, and density of states were also calculated. Derived from the electronic structure, valence X-ray photoelectron spectrum (vXPS) was simulated, matched well to the experimental one.

In chapter 5, "Physical Properties of Each PVDF Polymorphs", calculation of the normal mode vibration of form I--III crystals of PVDF was carried out at the level of PBE0/cc-pVTZ//PBE0/cc-pVTZ(d,p) to reproduce IR/Raman spectra. 
In the process, the author suspects the assignment of Raman scattering in the precedent study and suggests new assignment.
This also suggests that the forcefields of C--H stretching contained a mistake.
The amount of spontaneous polarization of three polar crystals (forms I, III and IV) was derived, and the author discussed about the long-range Coulomb interactions.
As a result, the amount of spontaneous polarization of form I PVDF yielded 176$\,$mC/m$^2$, and the amount of spontaneous polarization of form III and IV gave 71$\,$mC/m$^2$ and 85$\,$mC/m$^2$, respectively.

In chapter 6, "Conclusion", the results and the point of view are completed.
In this study, the  exhaustive calculation was performed carefully. 
For the first time, the amount of spontaneous polarization of three PVDF crystals are derived.

\onecolumn 

\twocolumn 
 
\onecolumn
\chapter{Theoretical Background} 


In this chapter, the computational methods employed through this study are described.

Now, the computational methods to be employed in quantum chemistry are divided into two category: molecular orbital theory and density functional theory.
Both of them are based on Sch\"odinger equation and are derived for nonempirical approximations and got the best energy as an eigenvalue and the set of orbitals as an eigenfunction through variational method.

In this section, behaviors of electrons in a crystal will be explained with Bloch's theorem,
and the summary of band theory will be described.

\section{Molecular orbital theory}

Molecular orbital (MO) theories expand an exact many-body wave function to a set of MOs and are Hartree-Fock (HF) method and post HF methods.
HF method  treats a single Slater determinant as ground state.
And post HF methods treat a linear combination of determinants which consist of a ground state determinant and excitation determinants.

\subsection{Hartree-Fock method}

HF method treats a wave function expressed in a single Slater determinant which contains many molecular orbitals.
A Slater determinant and its linear combinations adopt the characteristics of mathematical determinant to anti-symmetric principle of fermions.
A Slater determinant is described with molecular orbital $\chi_i(\vec{r}_j)$ as,
\begin{equation}
\|\chi_1\dots \chi_N\|=\frac{1}{N!}
\left|
\begin{array}{ccc}
\chi_1(\vec{r}_1) & \ldots & \chi_1(\vec{r}_N)\\
\vdots & \ddots & \vdots \\
\chi_n(\vec{r}_1) & \ldots & \chi_n(\vec{r}_N)
\end{array}
\right|.
\end{equation}
What the determinant will be zero when two molecular orbitals is same is the Fermi-Dirac statics.
And the asymmetric principle is equivalent to  what if the column or row were swapped, the sign of determinant change.
In this way, Slater determinant satisfy Pauli exclusion principle.

Hartree-Fock equation in a molecule, where the $A$-th atom has atomic number of $Z_A$, is described as:
\begin{eqnarray}
\epsilon_i \chi_i&=& \hat{F} \chi_i \\
\hat F_i &=& -\frac 12 \nabla ^2_i - \sum _A \frac{Z_A}{r_{iA}} + \sum_j (2\hat{J}_j+\hat{K}_j),\\
\hat J_j \chi_i(\vec r_i) &=& \int \frac{|\chi_j(\vec r_j)|^2}{\vec r_{ij}} \chi_i(\vec r_i) {\rm d}\vec r_j,\\
\hat K_j \chi_i(\vec r_i) &=& \int \frac{\chi^*_j(\vec r_j)\chi_i(\vec r_j)}{\vec r_{ij}}\chi_j(r_i) {\rm d}\vec r_j,
\end{eqnarray}
where $\epsilon$ is orbital energy, Fock operator $\hat{F}$ is an approximation of a one electron hamiltonian, Coulomb operator $\hat{J}$ gives classical Coulomb energy, and exchange operator $\hat{K}$ is derived from Pauli exclusion principle.

\subsection{Electronic correlations}
HF method treats the electrostatic energy of mean field of molecular orbital and gives higher energy than exact Schr"odinger eigenvalue.
The difference between Hartree-Fock and exact Schr\"odinger is known as electron correlation energy.
Post HF methods are employed to treat electronic correlations, 
and the methods approximate many-body wave function as linear combination of Slater determinants, a ground state and some excited states.
For example, configuration interaction (CI) and M{\o}ller-Plesset perturbation theory (MP) are well known methods.
To estimate electronic correlation precisely, it is needed treating so many excited determinants and consuming a large mount of computational resource and time.

\section{Density functional theory}
Density functional theory (DFT) is another quantum chemical method, is based on Hohenberg-Kohn theorems and represent total energy of electron as a functional of density of electron.
In DFT based on Kohn-Sham equation, exchange correlation energy appears the functional of the electron density.
DFT can evaluate electronic correlation more efficient than post HF methods.

\subsection{Thomas-Fermi model}
Thomas-Fermi model is a pioneer of DFT, and was a local density approximation (LDA) model of electron kinetic energy.

In 1927, Thormas proposed the form of LDA kinetic energy $T$ as a functional,\cite{2Thomas} using electron density $\rho(\vec r)$,
\begin{equation}
T^{\rm TF}=\frac{3}{10}(3\pi^2)^{2/3} \int \rho^{5/3}(\vec{r}){\rm d}\vec{r}.
\end{equation}
Then, Weizs\"acker derived the compensation term of TF model with the gradient of electron density,
\begin{equation}
T^{\rm W}=\frac{1}{8} \int \frac{|\nabla \rho|^{2}}{\rho} {\rm d} \vec{r},
\end{equation}
this kind of compensation is called generalized gradient approximation (GGA).\cite{2Weizsaecker}

These kinetic energy functional could not be reproduce chemical bonds, 
therefore Kohn-Sham equation, which represent kinetic energy as an operator, is widely used now.

\subsection{Kohn-Sham equation}
Kohn-Sham (KS) equation treats electron kinetic energy as operator of orbitals like HF method and treats exchange correlation energy as a functional of electron density.
Therefore the total KS energy is described as 
\begin{equation}
\begin{split}
E^{\rm KS} = \sum_i\left<\chi_i\left|-\frac 12\nabla^2\right|\chi_i\right> 
+\sum_A\int\frac{Z_A \rho}{\vec r-\vec r_A}{\rm d}\vec r 
+\int\int\frac{\rho(\vec r_i)\rho(\vec r_j)}{\vec r_{ij}}{\rm d}{\vec r_i}{\rm d}{\vec r_j} 
+E_{\rm xc}[\rho],
\end{split}
\end{equation}
where $E_{\rm xc}[\rho]$ is an exchange correlation functional described below.

\subsubsection*{LDA exchange functional} 
Slater derived that the LDA exchange functional can be described theoretically as following: \cite{2LDA}
\begin{equation}
E_{\rm x}^{\rm LDA}[\rho(\vec r)]=-\frac 34 \left(\frac 3\pi\right)^{1/3}\int \rho^{4/3} {\rm d}\vec r.
\end{equation}

\subsubsection*{GGA exchange functionals}
LDA exchange functional was theoretically derived for the inhomogeneous electron gas but has a large amount of error in real system.
GGA functional compensates this errors using gradient of electron density and was proposed many forms:
Becke (B88 or B), \cite{2B88}
Perdew-Wang (PW-GGA), \cite{2PW-GGA}
Perdew-Burke-Ernzerhof (PBE). \cite{2PBE}
For example, the formula of PBE exchange functional is simple and is described as
\begin{equation}
E_{\rm x}^{\rm PBE}= -\frac 34 \left(\frac 3\pi\right)^{1/3}\int \rho^{4/3} \left(1+\kappa-\frac{\kappa}{1+\frac{\mu}{\kappa} s^2}
\right) {\rm d}\vec r ,
\end{equation}
where $\kappa$ and $\mu$ are constants and 
\begin{equation}
s=\frac{|\nabla\rho|}{2(3\pi^2\rho)^{1/3}\rho}.
\end{equation}

\subsubsection*{Correlation functionals}
Correlation functionals are not derived in analytical like LDA exchange functional.

Vosko-Wilk-Nusair (VWN) functional is derived with fitting from the calculation with Monte Carlo method. \cite{2VWN}
There are some functionals:
LDA functionals, VWN, Perdew-Wang (PW-LDA); \cite{2PW-LDA}
GGA functionals, Lee-Yang-Parr (LYP), \cite{2LYP} Perdew-Burke-Ernzerhof. \cite{2PBE}
For example, the formula of PW-LDA is
\begin{equation}
E_{c}^{\rm PW-LDA}[\rho]=-2a\int d\vec r \rho(1-\alpha r_{\rm s})
\ln\left[1+\frac{1}{2a+\beta_1 r_{\rm s}^{1/2}+\beta_2 r_{\rm s}+\beta_3 r_{\rm s}^{3/2}+\beta_4 r_{\rm s}^{2}} \right]
\end{equation}
where $a$, $\alpha$, $\beta$s are constants.
\begin{equation}
\frac 43 r_{\rm s}^3=\frac 1\rho,
\end{equation}

\subsubsection*{Hybrid functionals}
Hybrid functional is a GGA functional compensated with HF exchange energy.
The compensation is done by mixing the HF energy and GGA exchange energy.

The popular B3LYP functional, Becke's three-parameter hybrid functional with LYP correlation, is:
\begin{equation}
E_{\rm xc}^{\rm B3LYP}=E_{\rm xc}^{\rm LDA}+0.20(E_{\rm x}^{\rm HF}-E_{\rm x}^{\rm LDA})+0.72(E_{\rm x}^{\rm GGA}-E_{\rm x}^{\rm LDA})+0.81(E_{\rm c}^{\rm GGA}-E_{\rm c}^{\rm LDA}),
\end{equation}
where $E_{\rm x}^{\rm HF}$ is HF exchange energy $\left<\phi\left|\hat{K}\right|\phi\right>$, 
suffix x and c means exchange and correlation, respectively. \cite{2B3LYP}
B88 exchange functional was employed as a GGA exchange functional, 
LYP correlation was chosen as a GGA correlation,
VWN correlation functional was employed as a LDA correlation functional.
The parameters are chosen empirically.

The other hybrid functional PBE0, Perdew-Burke-Ernzerhof zero-parameter, \cite{2PBE1, 2PBE0} hybrid functional is described simply as:
\begin{equation}
E_{\rm xc}^{\rm PBE0}=E_{\rm xc}^{\rm PBE}+\frac 14 \left(E_{\rm x}^{\rm HF}-E_{\rm x}^{\rm PBE}\right).
\end{equation}
The parameter 1/4 is derived theoretically.

\section{DFT and intermolecular force}
\begin{figure*}
\includegraphics[width=\textwidth]{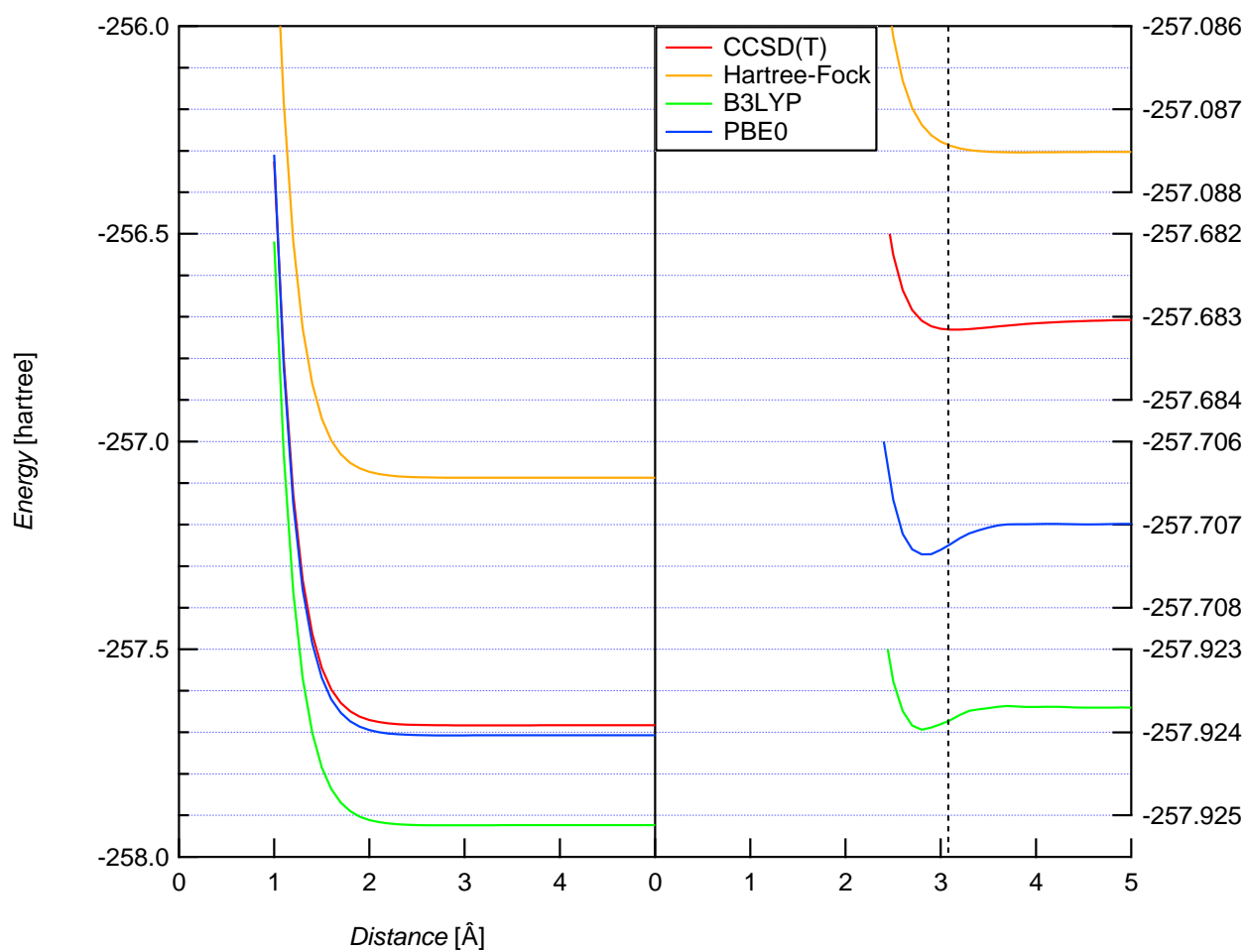}
\caption{Atomic distance and energy of neon dimer. The left char is overview. The right chart is magnified to show detail of each method. The vertical doted line is the van der Waals radius.}\label{2Neon}
\end{figure*}

\begin{figure*}
\includegraphics[width=\textwidth]{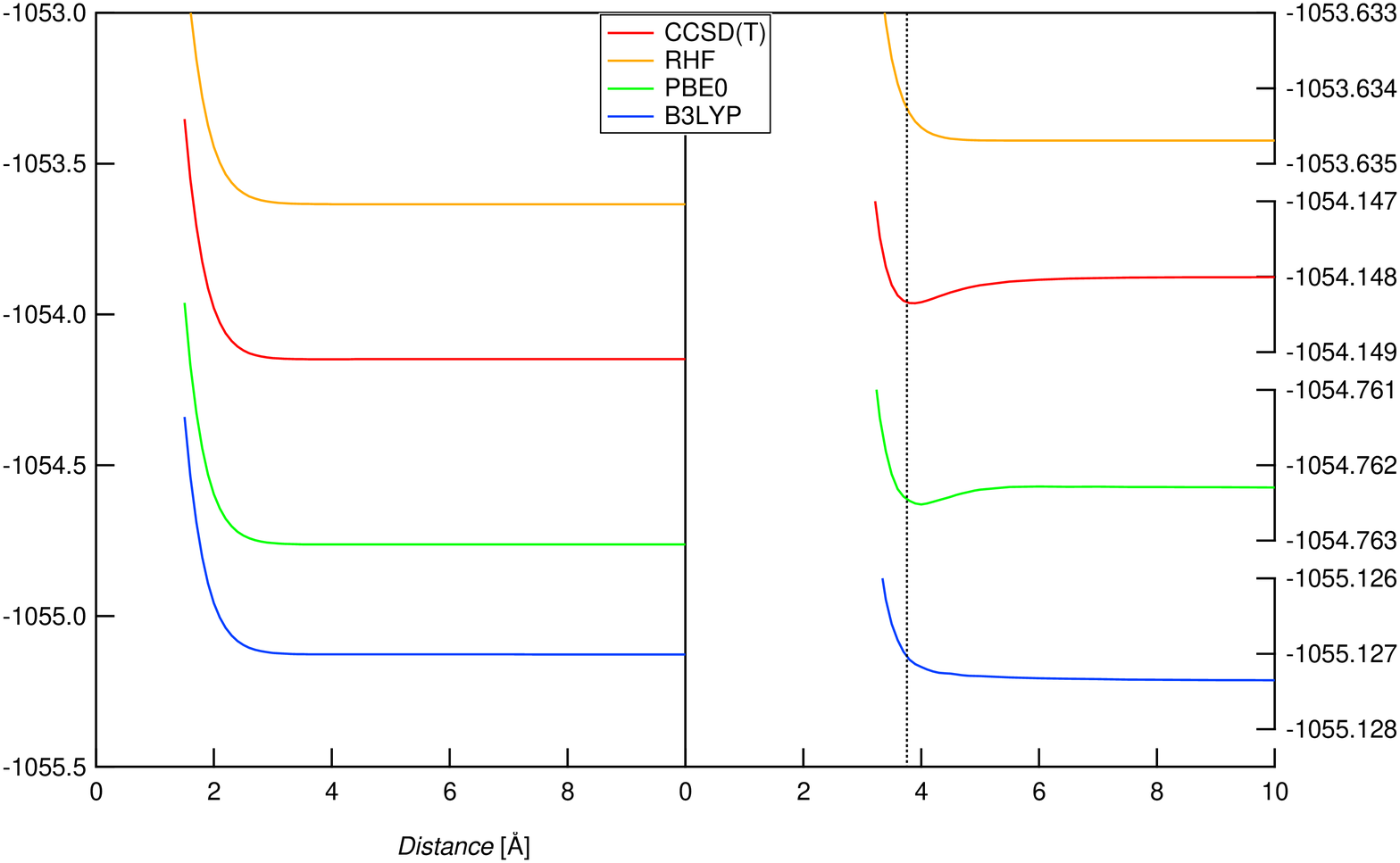}
\caption{Distance and energy of argon dimer.}\label{2Argon}
\end{figure*}

Quantum chemical methods were evaluated through the dependency between the energy and the distance for neon dimer (figure \ref{2Neon}) and argon dimer (Figure \ref{2Argon}).
Employed methods are followings:
post HF method, CCSD(T) with aug-cc-pVQZ basis set;
HF method;
hybrid functionals, B3LYP and PBE0 with cc-pVTZ.

\subsection*{Molecular orbital methods}
CCSD(T), post HF method, is the best approximation to exact Schr\"odinger equation in the methods and shows van der Waals force.
On the other hand, HF method do not show dispersion interaction for both neon and argon.

The London dispersion force, force between instantaneously induced dipoles, arises from electronic correlation, and HF method neglect the correlation.
So the energy difference between CCSD(T) and HF seems to be the exact correlation energy.

\subsection*{B3LYP}
About Ne$_2$, B3LYP reproduced the dispersion force.
However, about Ar$_2$, the energy decrease monotonically with increase in the distance.
It is said that B3LYP does not evaluate the intermolecular forces correctly.

\subsection*{PBE0}
There is dispersion force for both Ne$_2$ and Ar$_2$ with PBE0/cc-pVTZ.

Equivalent position is in good agreement with twice of van der Waals radius for Neon and is slightly far for argon.
The potential energy is slightly lower than CCSD(T).
For neon dimer, the curve is almost agrees in overview.
For argon dimer, the minimum is slightly deeper, 
and the deviation of energy from the exact is at nearly same level of correlation energy, about 1$\,$eV per one electron.

In this way, PBE0 could reproduce the molecular interaction approximately, of course it is not worth exact, and was applied for inorganic crystal.
In this study, it will be shown that the method can also be good approximation for organic polymer crystal.

\section{Basis sets}
In this study, the basis sets will be expressed as LCAO-GTO type.
LCAO is the method  molecular orbitals are expressed by the linear combination of atomic orbital like functions, s, p, d and f type functions, when AO like functions are arranged at the position of atoms.
With the GTO methods, an AO-like function is described in a linear combination of Gaussian type functions (GTF).

A GTF $\chi$ is expressed as a formula below:
\begin{eqnarray}
\chi(r;\alpha)&=&\sqrt{N}(x-X)^l (y-Y)^m (z-Z)^n \exp(-\alpha r^2)  \\
N&=&\frac{2^{2(l+m+n)+\frac{3}{2}}\alpha^\frac{3}{2}}{(2l-1)!!(2m-1)!!(2n-1)!!\pi^\frac{3}{2}},\nonumber
\end{eqnarray}
where $N$ is the normalization coefficient.
Then, the basis function is an approximation of atomic orbital and is described as a linear combination of GTFs as following:
\begin{eqnarray}
\phi(r)&=&\sum_i d_i \chi(r;\alpha_i),
\end{eqnarray}
the set of exponents $\alpha_i$ and coefficients $d_i$ is determined at each basis function of each element,
the combination is called a basis set.

Many basis sets are proposed.
In this study, the basis sets are treated with three categories.
The first include 6-31G$^{**}$ and 6-311G$^{**}$ developed by Pople and co-workers in early study of quantum chemistry and are widely used in the most in the field.
The second is developed by Godbout \textit{et al.} for the use with DFT calcuation and includes DZVP, DZVP2 and TZVP.
The last is correlation consistent polarized valence $N$-zeta basis set series developed by Dunning and includes cc-pVDZ and cc-pVTZ and is developed for accurate post HF calculations.
The three categories have the larger scale as the behind.

\begin{figure*}
\includegraphics[angle=-90, width=\textwidth]{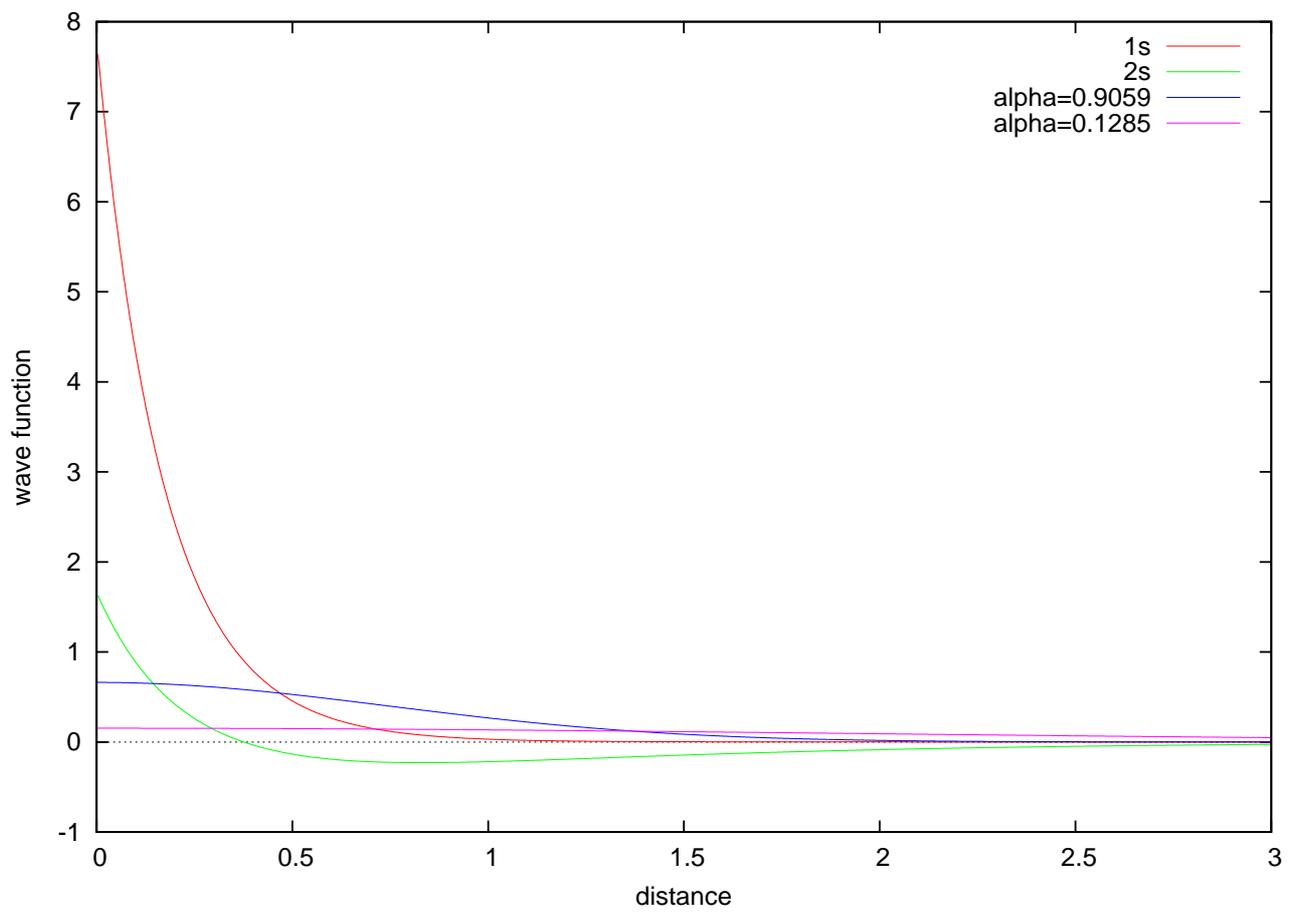}
\caption{Carbon s basis functions of cc-pVTZ basis set.}\label{CarbonS} 
\end{figure*}

\begin{figure*}
\includegraphics[angle=270, width=\textwidth]{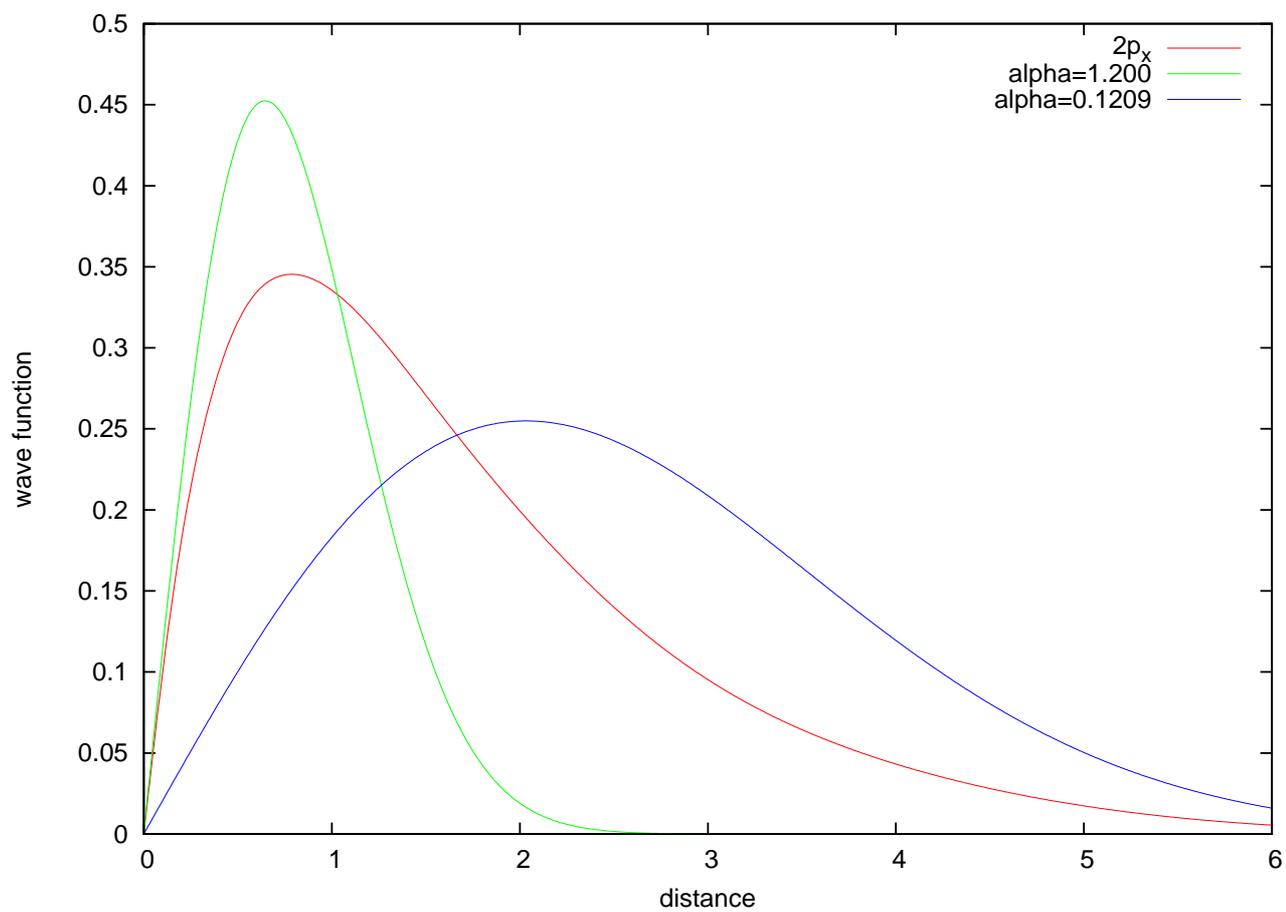}
\caption{Carbon p$\rm _x$ basis functions of cc-pVTZ basis set along x direction.}\label{CarbonP} 
\end{figure*}

For example, basis functions in cc-pVTZ basis set are illustrated in Figure \ref{CarbonS} for Carbon s functions and in Figure \ref{CarbonP} for s functions of Carbon and are described in Table \ref{CBasis}, \ref{FBasis} and \ref{HBasis} for carbon, fluorine and hydrogen, respectively.
In cc-pVTZ for carbon and fluorine, polarization functions are two d type function (d$^*$ and d$^{**}$) and one f type function (f$^*$), and each valence function (2s and 2p) is split in three functions to increase flexibility for variation.
For hydrogen, polarization functions are two p type functions (p$^{*}$, p$^{**}$) and one d type function (d$^*$), 
and valence functions are 1s, s' and s''.

\begin{table}
\begin{center}
\caption{cc-pVTZ basis set for carbon.}\label{CBasis}
\begin{tabular}{r@{}lrrrr}
\hline
\multicolumn{2}{c}{$\alpha$}&
\multicolumn{1}{c}{$d$(1s)}&
\multicolumn{1}{c}{$d$(2s)}&
\multicolumn{1}{c}{$d$(s')}&
\multicolumn{1}{c}{$d$(s'')}\\
\hline
8236.&&0.000531&$-0.000113$\\
1235.&&0.004108&$-0.000878$\\
280.&8&0.021087&$-0.004540$\\
79.&27&0.081853&$-0.018133$\\
25.&59&0.234817&$-0.055760$\\
8.&997&0.434401&$-0.126895$\\
3.&319&0.346129&$-0.170352$\\
0.&9059&0.039378&0.140382&1.000000\\
0.&3643&$-0.008983$&0.598684\\
0.&1285&0.002385&0.395389&&1.000000\\
\hline
\\
\cline{1-5}
\cline{1-5}
\multicolumn{2}{c}{$\alpha$}&
\multicolumn{1}{c}{$d$(2p)}&
\multicolumn{1}{c}{$d$(p')}&
\multicolumn{1}{c}{$d$(p'')}\\
\cline{1-5}
18.&71&0.014031\\
4.&133&0.086866\\
1.&200&0.290216&1.000000\\
0.&3827&0.501008\\
0.&1209&0.343406&&1.000000\\
\cline{1-5}
\cline{1-5}
\\
\cline{1-5}
\cline{1-5}
\multicolumn{2}{c}{$\alpha$}&
\multicolumn{1}{c}{$d$(d*)}&
\multicolumn{1}{c}{$d$(d**)}&
\multicolumn{1}{c}{$d$(f*)}\\
\cline{1-5}
1.&097&1.000000\\
0.&3180&&1.000000\\
0.&7610&&&1.000000\\
\cline{1-5}
\cline{1-5}
\end{tabular}
\end{center}
\end{table}

\begin{table}
\begin{center}
\caption{cc-pVTZ basis set for fluorine.}\label{FBasis}
\begin{tabular}{r@{}lrrrr}
\hline
\multicolumn{2}{c}{$\alpha$}&
\multicolumn{1}{c}{$d$(1s)}&
\multicolumn{1}{c}{$d$(2s)}&
\multicolumn{1}{c}{$d$(s')}&
\multicolumn{1}{c}{$d$(s'')}\\
\hline
19500.&&0.000507&$-0.000117$\\
2923.&&0.003923&$-0.000912$\\
664.&5&0.020200&$-0.004717$\\
187.&5&0.079010&$-0.019086$\\
60.&62&0.230439&$-0.059655$\\
21.&42&0.432872&$-0.140010$\\
7.&950&0.349964&$-0.176782$\\
2.&257&0.043233&0.171625&1.000000\\
0.&8815&$-0.007892$&0.605043\\
0.&3041&0.002384&0.369512&&1.000000\\
\hline
\\
\cline{1-5}
\multicolumn{2}{c}{$\alpha$}&
\multicolumn{1}{c}{$d$(2p)}&
\multicolumn{1}{c}{$d$(p')}&
\multicolumn{1}{c}{$d$(p'')}\\
\cline{1-5}
43.&88&0.016665\\
9.&926&0.104472\\
2.&930&0.317260&1.000000\\
0.&9132&0.487343\\
0.&2672&0.334604&&1.000000\\
\cline{1-5}
\\
\cline{1-5}
\multicolumn{2}{c}{$\alpha$}&
\multicolumn{1}{c}{$d$(d*)}&
\multicolumn{1}{c}{$d$(d**)}&
\multicolumn{1}{c}{$d$(f*)}\\
\cline{1-5}
3.&107&1.000000\\
0.&8550&&1.000000\\
1.&917&&&1.000000\\
\cline{1-5}
\end{tabular}
\end{center}
\end{table}

\begin{table}
\begin{center}
\caption{cc-pVTZ basis set of hydrogen.}\label{HBasis}
\begin{tabular}{r@{}lrrr}
\hline
\multicolumn{2}{c}{$\alpha$}&
\multicolumn{1}{c}{$d$(1s)}&
\multicolumn{1}{c}{$d$(s')}&
\multicolumn{1}{c}{$d$(s'')}\\
\hline
33.&87&0.006068\\
5.&095&0.045308\\
1.&159&0.202822\\
0.&3258&0.503903&1.000000\\
0.&1027&0.383421&&1.000000\\
\hline
\\
\hline
\multicolumn{2}{c}{$\alpha$}&
\multicolumn{1}{c}{$d$(p*)}&
\multicolumn{1}{c}{$d$(p**)}&
\multicolumn{1}{c}{$d$(d*)}\\
\hline
1.&407&1.000000\\
0.&3880&&1.000000\\
1.&057&&&1.000000\\
\hline
\end{tabular}
\end{center}
\end{table}

\section{Electron in solid}
Periodic boundary conditions are employed in the field of solid-state science and 
where potential $V$ is periodic with all lattice vector $\vec R$.
Then the electrons behave as Bloch wave.

The periodicity of potential can be described as:
\begin{equation}
V(\vec r)=V(\vec r +\vec R),
\end{equation}
where, $\vec R$ can be described with integer $N_i$ and the primitive lattice vector $\vec a_i$ as:
\begin{equation}
\vec R=\sum_{i=1}^3 N_i \vec{a}_i.
\end{equation}

Then, reciprocal vector $\vec k$ is introduced for convenience. 
A reciprocal vector of a real lattice vector has the form of Fourier transformation of original lattice, the relationship between real lattice vector and reciprocal lattice vector is $\vec R \cdot\vec k =2\pi n$, where $n$ is an integer.
A reciprocal vector $\vec k$ is written as: 
\begin{equation}
\vec k=\sum_{i=1}^3 \frac{m_i}{N_i}\vec b,
\end{equation}
where, $m_i$ is an integer, and the primitive reciprocal vectors $\vec b_i$ are reproduced as:
\begin{eqnarray}
\vec{b}_1&=&2\pi\frac{\vec{a}_2\times \vec{a}_3}{\vec{a}_1\cdot(\vec{a}_2\times\vec{a}_3)}, \\
\vec{b}_2&=&2\pi\frac{\vec{a}_3\times \vec{a}_1}{\vec{a}_1\cdot(\vec{a}_2\times\vec{a}_3)}, \\
\vec{b}_3&=&2\pi\frac{\vec{a}_1\times \vec{a}_2}{\vec{a}_1\cdot(\vec{a}_2\times\vec{a}_3)}. 
\end{eqnarray}
Among the reciprocal vectors, first Brillouin zone is the Wigner-Seitz cell of reciprocal lattices and is the primitive unit cell of the theory of electronic state in periodic potential.
If $\vec k$ were in infinite crystal ($L=\left| \sum n_i \vec k_i\right|$), $vec k$ are continuous.
However, the calculation will be performed under discrete $\vec k$.
Pack-Monkhorst Grid is the way to calculate solids with discrete $\vec k$.\cite{2Monkhorst}

Particles under PBC are behave in Bloch states: 
\begin{equation}
\psi_{n\vec k}=e^{i\vec k \vec r} \cdot u_{n\vec k}(\vec r),
\end{equation}
where, n is a band index. The function $u_{n\vec k}(\vec r)$ has a periodicity described as:
\begin{equation}
u_{n\vec k}(\vec r+\vec R)=u_{n\vec k}(\vec r).
\end{equation}
And $\psi_{n\vec k}$ changes the phase of it by translation alog $\vec R$:
\begin{equation}
\psi_{n\vec k}(\vec r+\vec R)=e^{i \vec k \vec R} \cdot \psi_{n\vec k}.
\end{equation}

Bloch theorem is mentioned above.
Here, the one electron Schr\"odinger equation is that:
\begin{equation}
\hat h \psi_{n \vec k}(\vec r)=\epsilon_{n \vec k} \psi_{n \vec k}(\vec r),
\end{equation}
where, $\epsilon_{n \vec k}$ is the eigenvalue and is continuous about $\vec k$,
and band structure derived from the dispersion of $\epsilon_{n \vec k}$ along $\vec k$.
Density of states (DOS) of electrons, $q(\epsilon)$ is the statistic of energy levels about all $\vec k$ in first Brilloin zone.
The states are $N(\epsilon+d\epsilon)$ in the energy range between $\epsilon$ to $\epsilon+d\epsilon$,
then,
\begin{equation}
q(\epsilon)=\frac{N(\epsilon+d\epsilon)}{d\epsilon}.
\end{equation}

\onecolumn 

\twocolumn

\chapter{Calculation of Form I PVDF}

In this chapter, to establish quantitative computational method, the lattice constants of PVDF Form I crystal were calculated in various method and were compared with the experimental values in precedent study using X-ray diffraction (XRD).
The agreement between the calculated result and the experimental value of the $c$-axis length along the chain direction is especially important to reproduce the mode chemical bondings.

\section{Introduction}

Form I crystal of PVDF is the ferroelectric crystal which has the largest spontaneous polarization among polymorphs, because the permanent dipole of all consistent units is packed in parallel with the $b$-axis direction. 
As a ferroelectric polymer, the crystal attracts attention of application; nonvolatile memory, the infrared sensor as the pyroelectricity material, some kind of sensors and the generator as the piezoelectric.

Kawai discovered that Form I crystal of PVDF is a polymer crystal with strong piezoelectricity. \cite{3Kawai}
Afterwards, Hasegawa \textit{et al.} determined that the crystal structure is $Cc2m$ orthorhombic system of all trans molecule. \cite{Hase72}
Furukawa observed polarization reversal at low temperature by applying high electric field. \cite{3Furukawa}
This experiment is the evidence of ferroelectricity.

To explain the electric property especially ferroelectricity, it is necessary to understand those; the molecular structure of consistent unit, intermolecular interactions, and the cooperative effect by the dipole-dipole interactions.
In particular, accurate molecular structure is need to evaluate the dipole moment,
$\vec{\mu}=\sum_i\,q_i\,\vec{r}_i$, where $q_i$ and $\vec{r}_i$ are the charge and the position of $i$-th atom, respectively.

\section{Purpose}

About Form I PVDF which has the most simple crystalline structure, a variety of quantum chemical methods will be adopted, the comparison of the lattice constants between the calculated and the experimental values will be performed, then the best method will be investigated.

As mentioned in Chapter 1, the density functional theory (DFT) is used to evaluate the electric properties of the material, recently.
However, in general, some methods with DFT do not reproduce dispersion interactions well; it is difficult to evaluate the lattice constants exactly.
So, the author will evaluate various methods and set the aim: the deviation less than 0.01$\,${\AA} of $c$ constant which reflects intra-molecular interactions and appears to depend on molecular structure, the best agreements of $a$ axis length which reflects simple intermolecular interaction and $b$ constant which reflects genuine intermolecular interaction and one of inter-dipole.

\section{Computational Details}

With the combination of Hamiltonian and basis set, the lattice constants of Form I PVDF were calculated through geometry optimization. 
All calculations for the primitive lattice of PVDF Form I were performed on CRYSTAL 06 \cite{3CRYSTAL}

For the Hamiltonians, Hartree-Fock method and some DFT methods were selected and shown below: 
a combination (SVWN) of LDA exchange functional and VWN corellation functional as the LDA functional,
two GGA functionals which are the set (BLYP) of B88 exchange functional \cite{B88} and LYP correlation functinal \cite{LYP} and PBE exchange-correlation functional \cite{PBE},
and two hybrid functional B3LYP \cite{B3LYP} and PBE0. \cite{PBE1, PBE0}
The details of the methods was described in chapter 2.

Next, the author examined the effect of basis sets:
series of basis set, 6-31G**, \cite{6-31G} 6-311G**, \cite{6-311G} developed by Pople and coworker;
two basis set used in DGauss, DZVP, DZVP2, TZVP; \cite{DZVP}
series of base set, cc-pVDZ, cc-pVTZ, developed by Dunning \cite{cc-pVxZ}.
cc-pVTZ could be treated with slight customization; 
for valence-triple zeta basis set about $p$-basis function, most and thirdly diffused gaussians are split from remaining. $s$-functions and polarization functions are keep original.

The geometry optimization is the process:
The initial structure had been given.
The electronic states in the structures was derived from variational principle.
The optimal geometry via variational principle was calculated.
Here, the initial structure is consisted from the set of the lattice constants and the geometries of atoms.
The electronic state, a set of orbitals and the electronic energy derived from the quantum mechanical technique; 
this electronic orbital calculation goes through the process of self consistent filed (SCF);
on the SCF calculation, trial functions which is component of wave function and made from a basis set and a set of the coefficients of basis functions is optimized for energy through these coefficients;
Through the SCF calculation, the best energy of initial structure is got;
then, the structure is changed to get the lowest energy via SCF calculation.
The result of geometry optimization is the structure, the total energy of structure and the sets of crystalline orbital and its energy.

\section{Results and discussion}
\subsection{Hamiltonian dependency}

\begin{figure*}
\begin{center}
\includegraphics[width=0.8\textwidth]{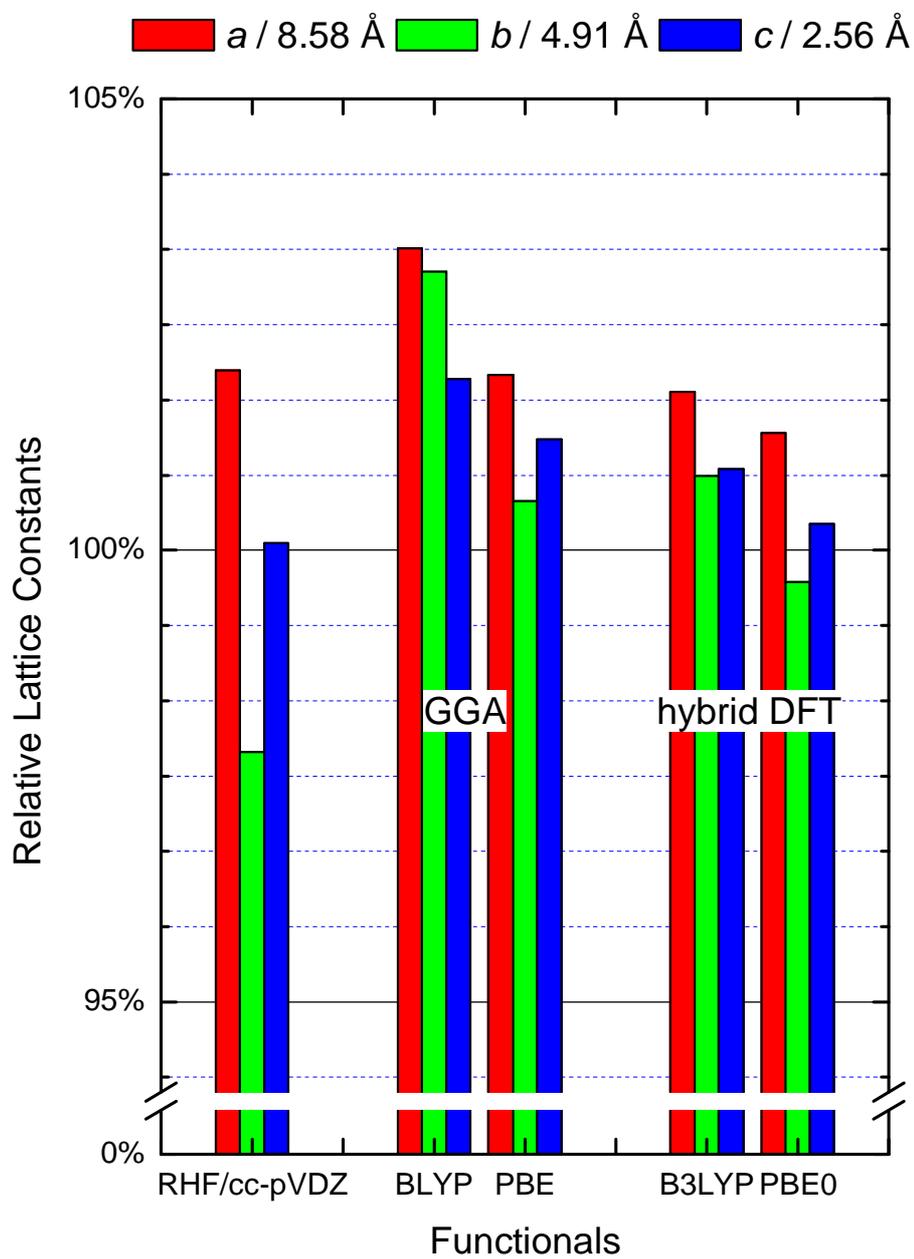}
\end{center}
\caption{
The Hamiltonian dependency of the set of lattice constants, relative to precedent study.
Basis set for DFT calculation is used cc-pVTZ;
the set for Hartree-fock (RHF) calculation is cc-pVDZ.
}\label{hamiltonians}
\end{figure*}

In this study, the largest basis set was cc-pVTZ.
In Figure \ref{hamiltonians}, the Hamiltonian dependency of the set of lattice constants is shown.
And, the lattice constant $c$ of this calculation is shown in Table \ref{c-length}

\begin{table*}
\caption{The calculated $c$ and its deviation from the reference value.}\label{c-length}
\begin{center}
\begin{tabular}{ccc}
\hline
Method&$c$ length [\AA]&Deviation [\AA]\\
\hline
Experimental value \cite{Hase72}&2.56\\
RHF/cc-pVDZ&2.562&$+0.00$\\
SVWN/cc-pVTZ&2.539&$-0.02$\\
BLYP/cc-pVTZ&2.609&$+0.05$\\
PBE/cc-pVTZ&2.591&$+0.03$\\
B3LYP/cc-pVTZ&2.582&$+0.03$\\
PBE0/cc-pVTZ&2.568&$+0.01$\\
\hline
\end{tabular}
\end{center}
\end{table*}

The calculated $c$ constant using HF/cc-pVDZ level, {2.562$\,$\AA} is agreed with experimental value well. 
For a problem of convergence, this calculation was performed with cc-pVDZ basis set.
It seems that the binding style of carbon backbone was evaluated exactly.
$a$ and $b$ slightly lack accuracy to neglect electronic correlation.

SVWN/cc-pVTZ gave higher energy than one of HF method; 
the exchange energy was too high.
The result of the lattice constants along $a$ and $b$ axis had deviations about $-10$\%; 
and $c$ was also underestimated.

BLYP/cc-pVTZ derives too low energy, it seems that the exchange energy was too low.
The all lattice constants were calculated greater than the experimental values considerably.
PBE/cc-pVTZ calculation also provided slightly greater constants.
These GGA functionals overestimate the chemical bondings greater from 101\%; 
it seems that GGAs are not suitable to evaluate the weak intermolecular interactions like van der Waals forces.

One of hybrid functional methods, B3LYP/cc-pVTZ was improved from the method with GGA functionals; 
the results of the lattice constants and energy are a little better.
$c$ was 0.02 angstroms greater from the reference value,
this method was inferior to PBE0.
The other hybrid functional method, PBE0/cc-pVTZ shows the best agreement about the result of lattice constants.

The characteristic of each Hamiltonian was analyzed each other. 
The HF and PBE0 methods showed in good agreement with the XRD value of the lattice constant $c$;
it can be thought that the chemical bonding style of carbon backbone was evaluated quantitatively.
In this system, many electrons crowded in fluorine atoms cause electronic correlation which the HF method neglects.
So, the evaluation of the behavior of electrons using the HF method has a problem.
In fact, the remaining lattice constants, $a$ and $b$ have some deviation from the adopted value; 
the total energy derived using the HF method is too high to be accepted.
On the other hand, the PBE0 method is hopeful.
The results in this system can be accepted.

\subsection{Basis set dependency}

\begin{figure*}
\includegraphics[width=\textwidth]{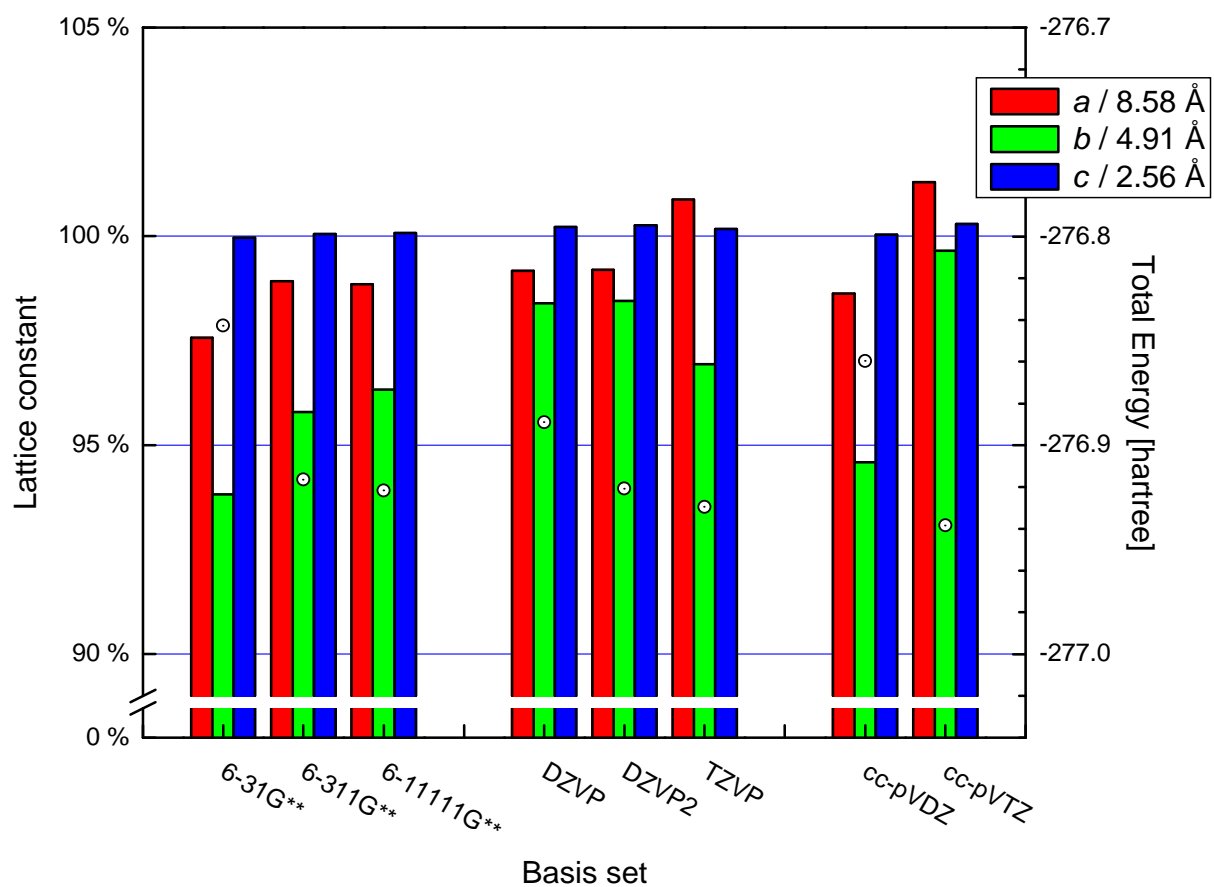}
\caption{The calculated lattice constants relative to experimental value using various basis sets.}\label{basissetdependency}
\end{figure*}

The basis set dependency on the energy and the lattice constants were explored with PBE0.
The relationship between each basis set and the set of lattice constants is shown in figure \ref{basissetdependency}.

$c$ of each calculation is matched in a deviation less than 1\%.
Also, $a$ axis length of each result was in agreement with the deviation less than 5\% from the reference; 
on the other hand, it seems that $b$ had large dependency to the basis set.

Two of the valence split double-zeta basis sets: 6-31$^{**}$ and cc-pVDZ gave relatively high total energy; 
and it seems that they are insufficient to calculate in a precision.
6-311G$^{**}$ was re-contracted to 6-11111G$^{**}$; the result of 6-11111$^{**}$ is not so enhanced from 6-311G$^{**}$.
The basis sets developed by Godbout \textit{et al.} which is used in DGauss, DZVP, DZVP2 and TZVP, indicated mathematical instability in this system; 
sometimes the calculation in certain conditions met mathematical catastrophe.
TZVP shows enhancement in energy in comparison with DZVP and DZVP2; 
but the agreement of $b$ with calculated and experimental values got worse.
The best basis set through this study is cc-pVTZ which is the largest one in this study; 
it gave the lowest energy and the best agreement.

All basis sets in this study were enough large to reproduce the chemical bonding style,
There is a balance along $b$-direction between the local filed (the dipole-dipole interactions) and other intermolecular interactions; it is difficult to evaluate $b$, quantitatively.
About the triple-zeta basis sets,
because the polarization functions are few in 6-311G$^{**}$ and TZVP, 
the balance between the dipole-dipole interactions and van der Waals interactions was not so good. 

\subsection{Population analysis}

Mulliken population analysis for the crystal with PBE0/cc-pVTZ was performed; 
the result is shown in table \ref{population}. 

\begin{table*}
\caption{
Popuation analysis for PVDF Form I crystal using PBE0/cc-pVTZ level calculation.
(a) Net atomic charge, where $n_{\rm e}$ and $Q_{\rm A}$ are the number of electrons and the atomic charge in each atom, respectively.
(b) Population assigned to each basis function.
The labels: 1s, 2s \textit{etc.} are the name of basis function.
}\label{population}
\begin{center}
\begin{tabular}{lcrr}
\multicolumn{4}{l}{(a) Atomic Charges.}\\
\cline{2-4}\noalign{\vspace{2pt}}
\cline{2-4} 
&Atom
&\multicolumn{1}{c}{$n_{\rm e}$}
&\multicolumn{1}{c}{$Q_{\rm A}$}\\
\cline{2-4}
&$\rm C_H$&5.681&$-0.481$\\
&$\rm C_F$&5.727&0.273\\
&H&0.711&0.289\\
&F&9.185&$-0.185$\\
\cline{2-4}\noalign{\vspace{2pt}}
\cline{2-4}
\end{tabular}
\begin{tabular}{lccccccccccc}
\multicolumn{12}{l}{(b) Populations}\\
\multicolumn{12}{l}{Carbon:}\\
\cline{2-12}\noalign{\vspace{2pt}}
\cline{2-12} 
&&1s&2s&s'&s''&2p&p'&p''&d*&d**&f*\\
\cline{2-12}
&$\rm C_H$&2.002&1.414&0.002&$-0.355$&3.651&$-0.041$&$-0.281$&0.017&0.068&0.004\\
&$\rm C_F$&1.998&1.358&0.016&$-0.358$&2.833&$-0.010$&$-0.465$&0.052&0.274&0.030\\
\cline{2-12}\noalign{\vspace{2pt}}
\cline{2-12} 
\\
\multicolumn{8}{l}{Hydrogen:}\\
\cline{2-8}\noalign{\vspace{2pt}}
\cline{2-8} 
&&1s&s'&s''&p*&p**&d*\\
\cline{2-8}
&H&1.379&$-0.388$&$-0.315$&0.008&0.024&0.002\\
\cline{2-8}\noalign{\vspace{2pt}}
\cline{2-8} 
\\
\multicolumn{11}{l}{Fluorine:} \\
\cline{2-12}\noalign{\vspace{2pt}}
\cline{2-12} 
&&1s&2s&s'&s''&2p&p'&p''&d*&d**&f*\\
\cline{2-12}
&F&1.990&1.885&0.016&0.013&5.394&$-0.091$&$-0.036$&0.001&0.013&0.001\\
\cline{2-12}\noalign{\vspace{2pt}}
\cline{2-12} 
\end{tabular}
\end{center}
\end{table*}

Fluorine atoms had a negative charge; 
hydrogen atoms had a positive charge;
the charge of carbon atoms bonded to hydrogen were negative; 
carbon atoms bonded to fluorine were positive.

It seems that the inner core electrons behaved mostly same as electrons in the isolated atoms; 
of course, the orbital energies were changed.
The valence electrons in carbon formed sp$^3$-like manner and might tighten.
A number of electrons populated the polarization functions of carbon;
the electrons populated in the d$^{**}$ function in the $\rm C_F$ atom was especially large amount. 
The electrons in hydrogen might be drawn to the nucleus.
The fluorine electrons in the crystal might be very similar to the ones in the isolated atoms.

Through the Mulliken population analysis, the electrons in each atom in the crystal were examined.
It was discovered that the effect by the fluorine atoms distort the atomic orbitals in the $\rm C_F$ atoms.

\section{Conclusion}

The calculation for PVDF form I crystal employed PBE0/cc-pVTZ level seems to be suitable to evaluate the crystals of PVDF.

The calculated lattice constants were in good agreement with a past study: 
$a=8.691\,$\AA (the reference value is 8.58$\,$\AA), 
$a=4.892\,$\AA (the reference value is 4.91$\,$\AA), 
$a=2.568\,$\AA (the reference value is 2.56$\,$\AA).
The deviation of $a$ from the reference to this calculation is +1.3\%; 
$-0.3$\% of the $b$ constant and +0.3\% of $c$ are calculated.
In the calculation, the total energy is $-276.9385\,$hartrees.
This results seem valid.

The basis set dependency and the population analysis lead the conclusion that a number of fine polarization functions are important to evaluate the weak interactions in this system in quantitative.
For example, the polarization functions in cc-pVTZ are diffused enough and are 2df for C and F and 2pd for H.

The calculation at the PBE0/cc-pVTZ level seems to be suitable to evaluate the crystals of PVDF.

\onecolumn

\twocolumn

\chapter{Crystalline and Electronic Structure of Polymorphs of PVDF}

In this chapter, the calculation employing the method PBE0/cc-pVTZ, which gave the best agreement with past study in previous chapter, will be performed in each polymorph of PVDF.
To examine the precision of the method, reproduction of the valence X-ray photoelectron spectrum is aimed at.

\section{Introduction}

PVDF has at least four crystalline polymorphs which are referred to as Forms I, III, IV and II or $\beta$, $\gamma$, $\delta$ and $\alpha$ phases, respectively. In 1972, Hesegawa \textit{et al.} decided the crystal structure of form I and II crystals of PVDF. \cite{4Hase72}
In 1980, Bachmann \textit{et al.} observed the crystal structure of form IV. \cite{4delta} 
In 1981, Lovinger reported the adopted structure of form III. \cite{4gamma} 

Each crystal consists of one of the three manners of molecular conformation, 
have different space group, 
indicates \textit{parallel} or \textit{antiparallel} orientation in the direction perpendicular to the molecular chain direction.
and indicate another orientation, "up-up" or "up-down" along the chain direction. 
Form I crystal consists of all trans (TT) molecules.
The trans-gauche-trans-minus gauche ($\rm TGT\bar{G}$) chains are observed in the crystalline forms II and IV.
Form III crystal consists of an intermediate conformation of $\rm T_3GT_3\bar{G}$.
Form I crystal has the crystal system of a base-centered orthorhombic ($oC$) one.
Form III crystal is in a base-centered monoclinic ($mC$) system.
Form IIV crystal and II are in a primitive orthorhombic ($oP$) one and in a primitive monoclinic one, respectively.
Molecular packing in crystalline forms I, III and IV is a parallel manner;
these three polymorphs are polar crystals and are regarded as ferroelectrics.
Form II crystal is non-polar, the chains are aligned antiparallel.
According to the past studies, the orientation of up-up or up-down is statically random in crystalline forms III, IV and II. In this study, the crystal structure of each polymorph is regarded as the one in Table \ref{structure} for simplicity.

\begin{table*}
\caption{The crystallographic system and chain alignment of PVDF polymorphs. 
Bravis lattice and space group of each crystals are described in Pearson symbol and Herrman-Morgan notation, respectively. 
Conformation is written in common symbol in polymer science. 
The symbol of alignment is in reference \cite{4KG}.}\label{structure}
\begin{center}
\begin{tabular}{p{10em}p{5em}p{5em}p{5em}p{5em}}
\hline
&
\multicolumn{1}{c}{Form I}&
\multicolumn{1}{c}{Form III}&
\multicolumn{1}{c}{Form IV}&
\multicolumn{1}{c}{Form II}\\
\hline
Bravis lattice&
\multicolumn{1}{c}{$oC$}&
\multicolumn{1}{c}{$mC$}&
\multicolumn{1}{c}{$oP$}&
\multicolumn{1}{c}{$mP$}\\
Space group&
\multicolumn{1}{c}{$Cc2m$}&
\multicolumn{1}{c}{$Cc$}&
\multicolumn{1}{c}{$P2_1cn$}&
\multicolumn{1}{c}{$P2_1/c$}\\
\hline
Conformation&
\multicolumn{1}{c}{TT}&
\multicolumn{1}{c}{$\rm T_3GT_3\bar{G}$}&
\multicolumn{1}{c}{$\rm TGT\bar{G}$}&
\multicolumn{1}{c}{$\rm TGT\bar{G}$}\\
Allignment&
\multicolumn{1}{c}{parallel}&
\multicolumn{1}{c}{parallel}&
\multicolumn{1}{c}{parallel}&
\multicolumn{1}{c}{antiparallel}\\
&&
\multicolumn{1}{c}{up-up}&
\multicolumn{1}{c}{up-down}&
\multicolumn{1}{c}{up-down}\\
\hline
\end{tabular}
\end{center}
\end{table*}

\section{Computational details}

\begin{figure}
\begin{center}
\includegraphics{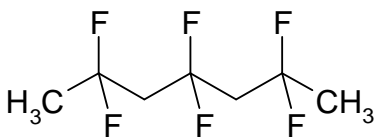}
\end{center}
\caption{The strucuture of an oligomer: 2,2,4,4,6,6-hexafluoroheptane.}\label{oligomer}
\end{figure}

The lattice constants and atomic positions of Form I--IV of PVDF with the three dimensional Bloch periodic boundary condition (3D-PBC) using primitive cells were calculated by geometry optimization with CRYSTAL09, \cite{4CRYSTAL} and were compared with the experimental adopted values.
To know the correct molecular structure, 2,2,4,4,6,6-hexafluoroheptane (figure \ref{oligomer}) is also calculated using MP2/aug-cc-pVTZ with Gaussian 03 program.\cite{4GAUSSIAN} 
With the molecular and electronic structure, the band structure and the density of states (DOS) were derived; With the results, vXPS was simulated.

\section{Results and discussion}

\subsection{Packing energy}

Total energy per $\rm CH_2CF_2$ unit minimized by optimization is 
-276.9385$\,$hartrees for Form I, 
-276.9839$\,$hartrees for Form III, 
-276.9394$\,$hartrees for Form IV, 
-276.9396$\,$hartrees for Form II.
These energy values include the formation energy of atoms from nuclei and electrons and that of polymer molecules from these atoms. 
The relative energy to Form I is 
$-2.7\,$kJ/mol for Form III,
$-2.5\,$kJ/mol for Form IV,
$-2.9\,$kJ/mol for Form II, respectively
Few values were indicated.

According to the basic chemical equilibrium theory, the equilibrium constant $K$ can be lead from the equation: 
\begin{equation}
\Delta G=-RT\ln{K},
\end{equation}
where $\Delta G$, $R$ and $T$ are the change in Gibbs free energy, the gas constant and the temperature, respectively.
The energy of \textit{ca.} 3$\,$kJ/mols leads 3.4 of the equilibrium constant. 
Of course, the energy is not a Gibbs energy.
And this problem is not a chemical equilibrium.
Even so, the difference is small enough.

This very small difference of energy may be one of the explanation of the stability of all polymorph.

\subsection{Crystal structure} 

\begin{table*}
\caption{The lattice constants (Angstroms for $a$, $b$ and $c$. Degrees for $\beta$) of PVDF polymorphs.  Calculated value (reference value) and the relative deviation are shown in each column.}\label{lattice constants}
\begin{center}
\begin{tabular}{cr@{ }r@{ }rr@{ }r@{ }rr@{ }r@{ }rr@{ }r@{ }r}
\hline
&
\multicolumn{3}{c}{Form I}&
\multicolumn{3}{c}{Form III}&
\multicolumn{3}{c}{Form IV}&
\multicolumn{3}{c}{Form II}\\
\hline
&
\multicolumn{3}{c}{Orthorhombic}&
\multicolumn{3}{c}{Monoclinic}&
\multicolumn{3}{c}{Orthorhombic}&
\multicolumn{3}{c}{Monoclinic}\\
&
\multicolumn{3}{c}{$Cm2m$}&
\multicolumn{3}{c}{$Cc$}&
\multicolumn{3}{c}{$P2_1cn$}&
\multicolumn{3}{c}{$P2_1/c$}\\
\hline
$a$&
8.691&(8.58)&{$+1.3\%$}&
4.892&(4.96)&{$-1.4\%$}&
5.011&(4.96)&{$+1.0\%$}&
5.029&(4.96)&{$+1.3\%$}\\
$b$&
4.893&(4.91)&{$-0.4\%$}&
9.568&(9.67)&{$-1.1\%$}&
9.999&(9.64)&{$+3.7\%$}&
9.976&(9.64)&{$+3.5\%$}\\
$c$&
2.568&(2.56)&{$+0.3\%$}&
9.226&(9.20)&{$+0.3\%$}&
4.645&(4.62)&{$+0.5\%$}&
4.646&(4.62)&{$+0.6\%$}\\
$\beta$&
&&&
94.06&(93.0)&&
&&&
90.40&(90.0)\\
&&&&
\multicolumn{3}{r}{+18mrad}&
&&&
\multicolumn{3}{r}{+7mrad}\\
\hline
\end{tabular}
\end{center}
\end{table*}

Calculated lattice constants are listed and compared with experimental values from references \cite{4Hase72, 4gamma, 4delta, 4Hase72} in Table \ref{lattice constants}.
In this calculation method, the lattice constants can be determined because the dispersion force is appropriately expressed.
The lattice constant, $c$, along the chain direction of each crystal form agrees well with the corresponding experiment value within 1\% of deviation.
In addition, $a$ and $b$ also agree fairly well even though they are directions that they are not connected by valence bonds, and are generally difficult to determine by calculation.
Thus, the calculated structure can be regarded as precise.

The atomic positions of each polymorph are shown in Table \ref{ACI}--\ref{ACII}

\begin{table}
\caption{Calculated atomic coordinates of Form I PVDF.}\label{ACI}
\begin{center}
$a$ = 8.6915 \AA, 
$b$ = 4.8929 \AA, 
$c$ = 2.5675 \AA\\[0.5pc]
\begin{tabular}{crrr}
\hline
\multicolumn{1}{c}{Atoms}&
\multicolumn{1}{c}{$x/a$}&
\multicolumn{1}{c}{$y/b$}&
\multicolumn{1}{c}{$z/c$}\\ 
\hline
C&0.00000&0.00000&0.00000\\
C&0.00000&0.16922&0.50000\\
F&0.12518&0.33861&0.50000\\
H&0.10140&$-0.13091$& 0.00000\\
\hline
\end{tabular}
\end{center}
\end{table}

\begin{table}
\caption{Calculated atomic coordinates of Form III PVDF.}\label{ACIII}
\begin{center}
$a$ = 4.8923 \AA, 
$b$ = 9.5677 \AA, 
$c$ = 9.2259 \AA, \\
$\beta$ = 94.0618$^\circ$\\[0.5pc]
\begin{tabular}{crrr}
\hline
\multicolumn{1}{c}{Atoms}&
\multicolumn{1}{c}{$x/a$}&
\multicolumn{1}{c}{$y/b$}&
\multicolumn{1}{c}{$z/c$}\\ 
\hline
C&0.05221&0.01228 &  0.00573\\
C&0.14279&$-0.11365$&$-0.07845$\\
C&0.04666&$-0.12082$&$-0.23867$\\
C&0.13111&$-0.00037$&$-0.33272$\\
F&0.16399&0.13109&$-0.04893$\\
F&$-0.22582$&0.02980&$-0.01751$\\
F&0.14793&$-0.24434$&$-0.28910$\\
F&$-0.23256$&$-0.13677$&$-0.25221$\\
H&0.06739&$-0.20835$&$-0.02863$\\
H&0.36628&$-0.11754$&$-0.06936$\\
H&0.04106&0.09536&$-0.29321$\\
H&0.35336&0.01087&$-0.31781$\\
\hline
\end{tabular}
\end{center}
\end{table}

\begin{table}
\caption{Calculated atomic coordinates of Form IV PVDF.}\label{ACIV}
\begin{center}
$a$ = 4.6008 \AA, 
$b$ = 8.7785 \AA, 
$c$ = 4.5582 \AA\\[0.5pc]
\begin{tabular}{crrr}
\hline
\multicolumn{1}{c}{Atoms}&
\multicolumn{1}{c}{$x/a$}&
\multicolumn{1}{c}{$y/b$}&
\multicolumn{1}{c}{$z/c$}\\ 
\hline
C&0.23015&   0.18237&  $-0.15136$\\
C&     0.33130&   0.18466&   0.16506\\
F&     0.23331&   0.05079&   0.28839\\
F&    $-0.37093$&   0.17527&   0.17682\\
H&     0.31052&   0.07723&  $-0.24980$\\
H&    $-0.00735$&   0.17932&  $-0.15434$\\
\hline
\end{tabular}
\end{center}
\end{table}

\begin{table}
\caption{Calculated atomic coordinates of Form II PVDF.}\label{ACII}
\begin{center}
$a$ = 5.0294 \AA, 
$b$ = 9.9762 \AA, 
$c$ = 4.6462 \AA, \\
$\beta$ = 90.4027$^\circ$\\[0.5pc]
\begin{tabular}{crrr}
\hline
\multicolumn{1}{c}{Atoms}&
\multicolumn{1}{c}{$x/a$}&
\multicolumn{1}{c}{$y/b$}&
\multicolumn{1}{c}{$z/c$}\\ 
\hline
C&0.23270&0.19295&$-0.14389$\\
C&0.32210&0.19084&0.16994\\
F&0.22361&0.07318&0.27958\\
F&$-0.40758$&0.17731&0.18435\\
H&0.30505&0.10007&$-0.23812$\\
H&0.01615&0.18939&$-0.14954$\\
\hline
\end{tabular}
\end{center}
\end{table}

\subsection{Molecular structure}

The calculated bonding parameters of oligomer, 2,2,4,4,6,6-hexafluoroheptane, employing MP2/aug-cc-pVTZ and the parameters of Form I crystal using PBE0/cc-pVTZ are shown in Table \ref{bonding}
In this calculation of oligomer, the torsion angle was fixed to 180$^\circ$.
The calculated internal rotation angle in Form I crystal is 180$^\circ$.

\begin{table}
\begin{center}
\caption{The bonding parameters of oligomer and Form I PVDF.}\label{bonding}
\begin{tabular}{ccc}
\hline
&Oligomer&Crystal\\
\hline
C-C distance [\AA]&1.525&1.528\\
C-F distance [\AA]&1.363&1.368\\
C-H distance [\AA]&1.090&1.090\\
C-C-C angle [deg.]&114.4&114.4\\
F-C-F angle [deg.]&107.0&105.4\\
H-C-H angle [deg.]&108.7&108.0\\
\hline
\end{tabular}
\end{center}
\end{table}

The molecular structure is decided because of the agreement between the result of calculation of Form I crystal using hybrid DFT and the calculation of oligomer with accurate molecular orbital method.
Additionally, the gauche internal rotations of Form III, IV and II are 59.2$^\circ$, 61.2$^\circ$ and 61.9$^\circ$, respectively.

The H-C-H angle is almost the tetrahedral angle of 109.5$^\circ$, and the C-H distance is nearly equal to the typical length of alkane, 1.09$\,$\AA.
The lattice constant $c$ and the angle of C-C-C are slightly, but significantly, largee than those of polyethylene (2.534$\,$\AA and 112$^\circ$). Thei may be caused by the repulsion between fluorine atoms that are bonded to different carbon atoms.
The C-F bond is longer than its typical value (1.34$\,$\AA)and the angle of F-C-F is smaller than that of tetrahedral angle. The distance between F atoms bonded to the same carbon atom is less than twice the van der Walls raius of a fluorine of 1.35$\,$\AA (otherwise 1.33$\,$\AA).
This suggests that the two fluorine atoms have some attractive interaction, which will be discussed further later.
The typical C-F bond length is 1.34$\,$\AA and is less than a sum of the covalent radius of carbon (0.77) and the radius of fluorine (0.64); 
this shortening is known as the resonance between ionic and covalent bonding.
In this case, the C-F distance ias

\subsection{Electronic band structure}

\begin{figure*}
(d)\\
\includegraphics[width=\textwidth]{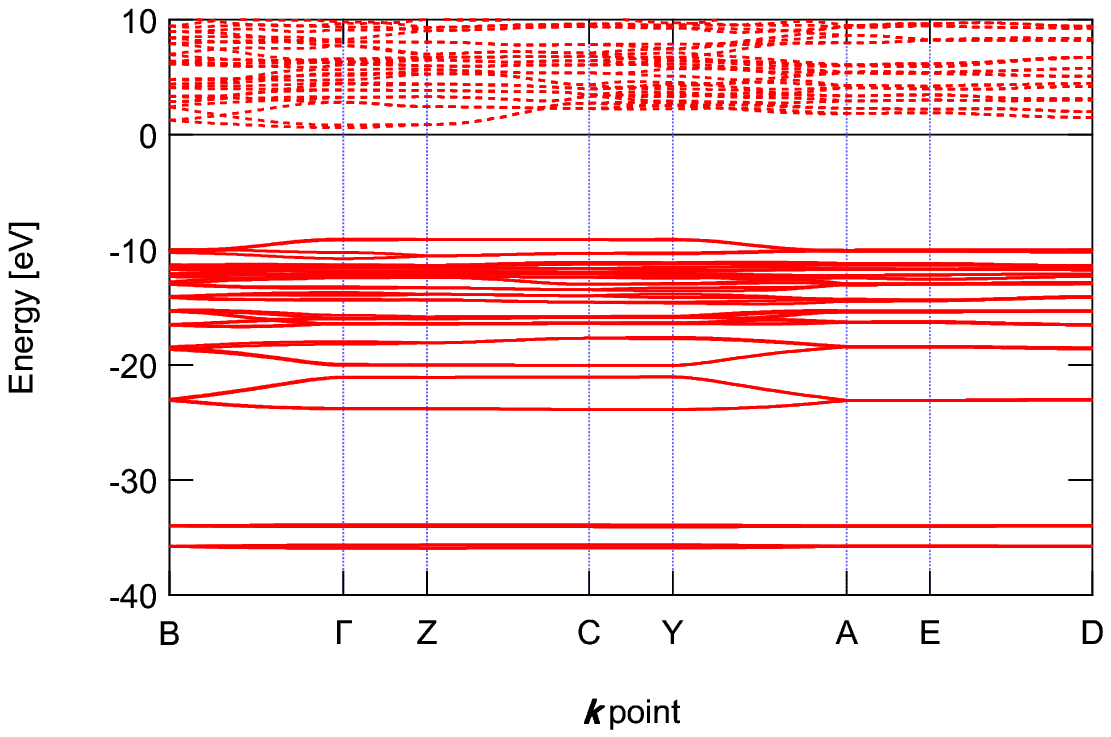}\\
\includegraphics{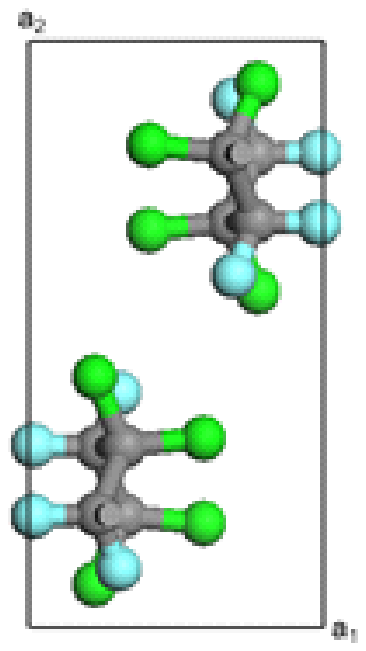}
\includegraphics{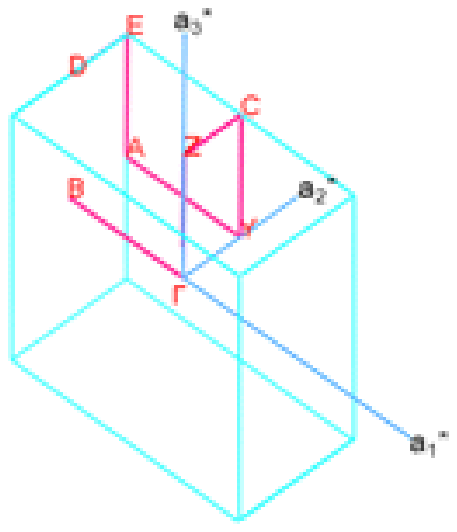}\\

\caption{Band structures of polymorphs of PVDF 
(a) Form I, (b)Form III, (c) Form IV and (d) Form II.
The crystallographic axes in primitive cells and the position of $k$-points in reciprocal lattices are taken as indicated.
For every form, $a_3$ axis is the fiber axis and the direction is taken as crystallograpic axes form the right-handed system}\label{Band} 
\end{figure*}

In Figure \ref{Band}, the calculated electronic band structure is shown.
Form I crystal has one CH$_2$CF$_2$ unit in primitive cell;
16 electron bands exist in the crystal.
The remaining polymorphs have the 4 units and 64 electron bands in each primitive cell.
The 1s orbital level of C and F is too low to be shown in the diagrams.

\begin{figure*}
(a)\\
\includegraphics[width=\textwidth]{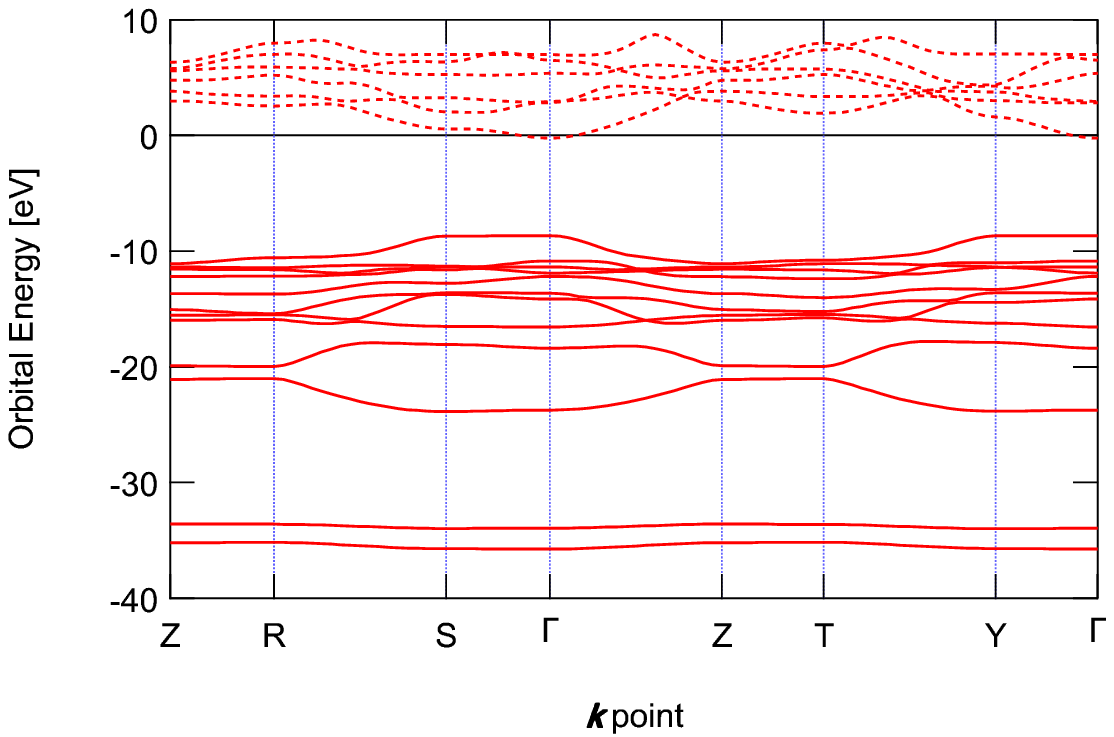}\\
\includegraphics{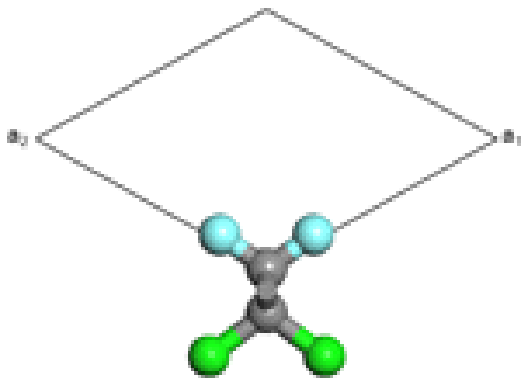}
\includegraphics{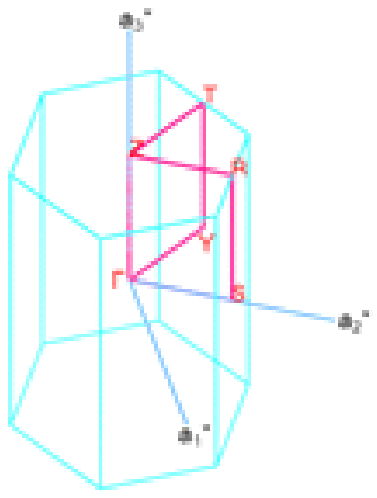}
\end{figure*}
\begin{figure*}
(b)\\
\includegraphics[width=\textwidth]{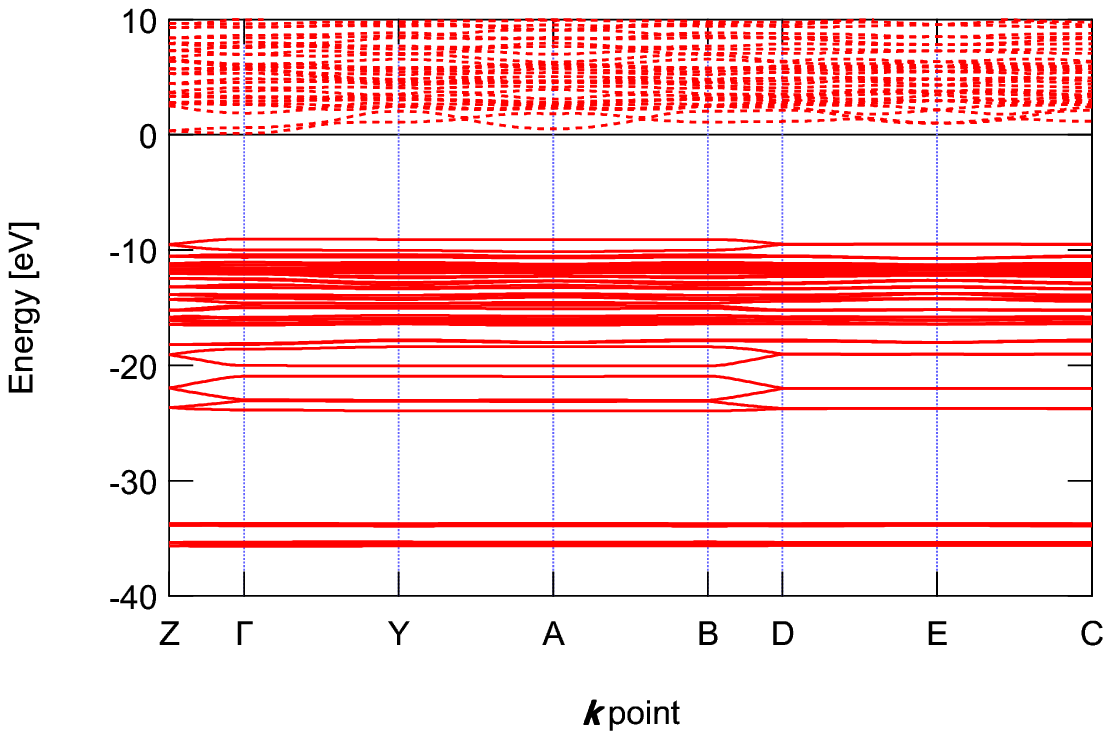}\\
\includegraphics{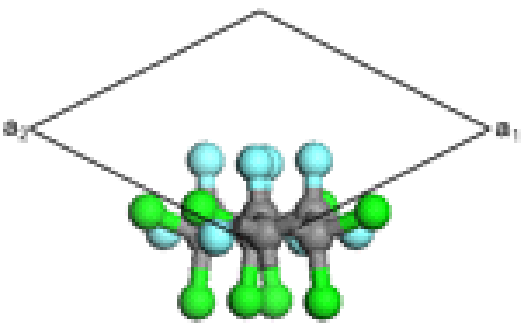}
\includegraphics{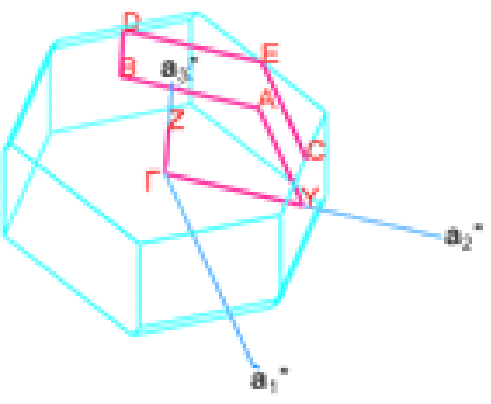}
\end{figure*}
\begin{figure*}
(c)\\
\includegraphics[width=\textwidth]{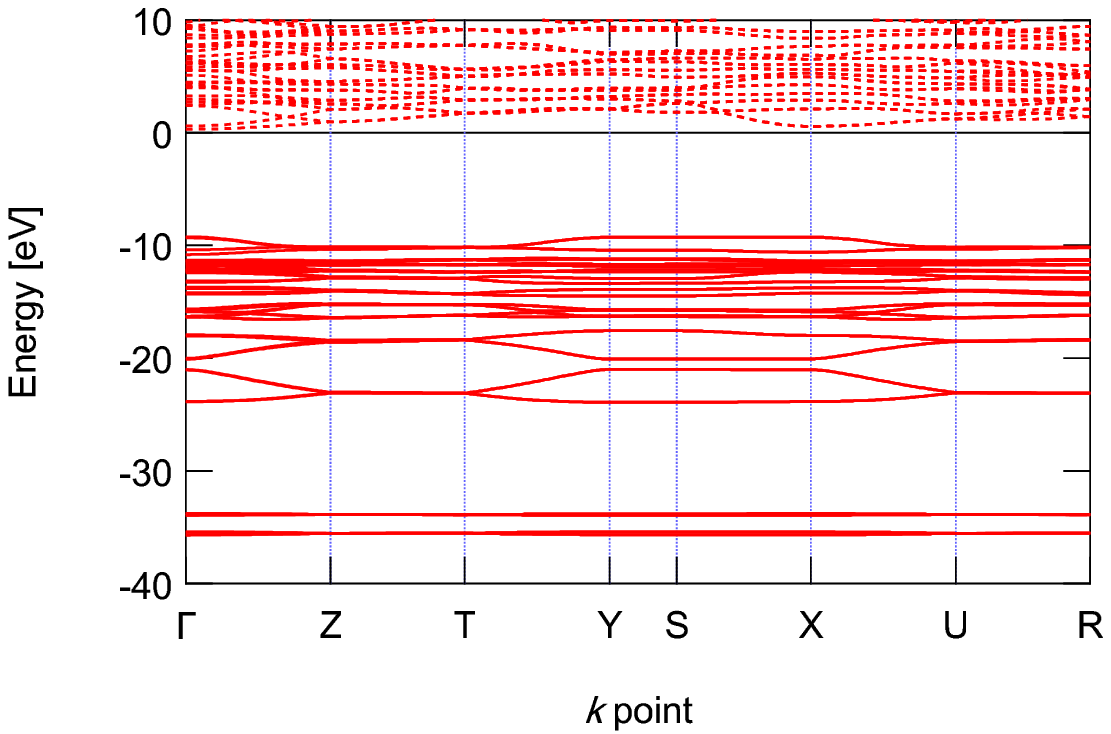}\\
\includegraphics{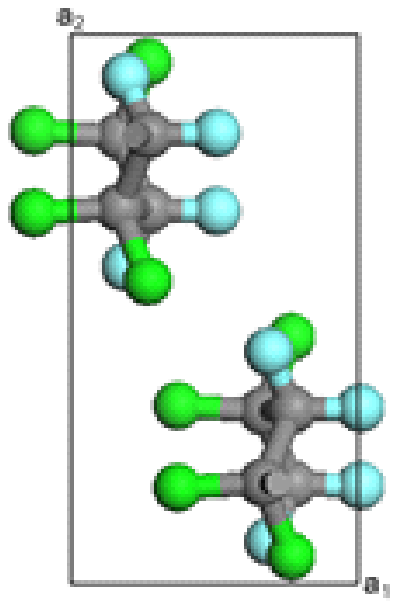}
\includegraphics{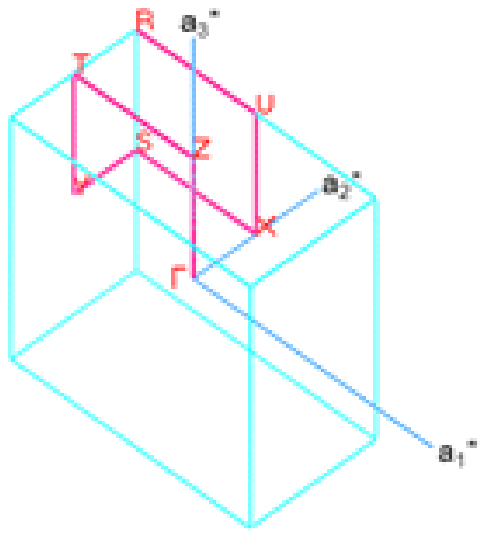}
\end{figure*}

In the band diagram of Form I, the first conduction band has lower energy than the vacuum level,
and is shaped like a quadratic function incidentally.
the crystal is possible to behave as a wide bandgap \textit{n}-type semiconductor.
Because the three crystals have a degenerated conduction band above the vacuum level, they are hardly thought to have such a semiconductivity.

At about -35$\,$eV, two split electron bands can be seen that are 2s electrons of fluorine.
The 2$\,$eV gap provides further evidence for a weak attractive interaction between the two fluorine atoms in a CF$_2$ unit; in other words, the fluorine atoms have a weak chemical bond and thereby the C-F bonds are wekened. Two groups of dipersed bands can be seen between $-24\,$eV, which are $\sigma$ bands, and $-21\,$eV, and between $-20\,$eV and $-18\,$eV, which are $\pi_{\rm b}$ band, respectively. Because thes orbitals involve C-C bonds and are distributed along the $c$-axis, corresponding bands are dipersed along $c^*$ direction, although they are degenerated along other directions.

\subsection{Density of states and simulation of valence X-ray photoelectron spectrum} 

\begin{figure*}
\begin{center}
\includegraphics[width=\textwidth]{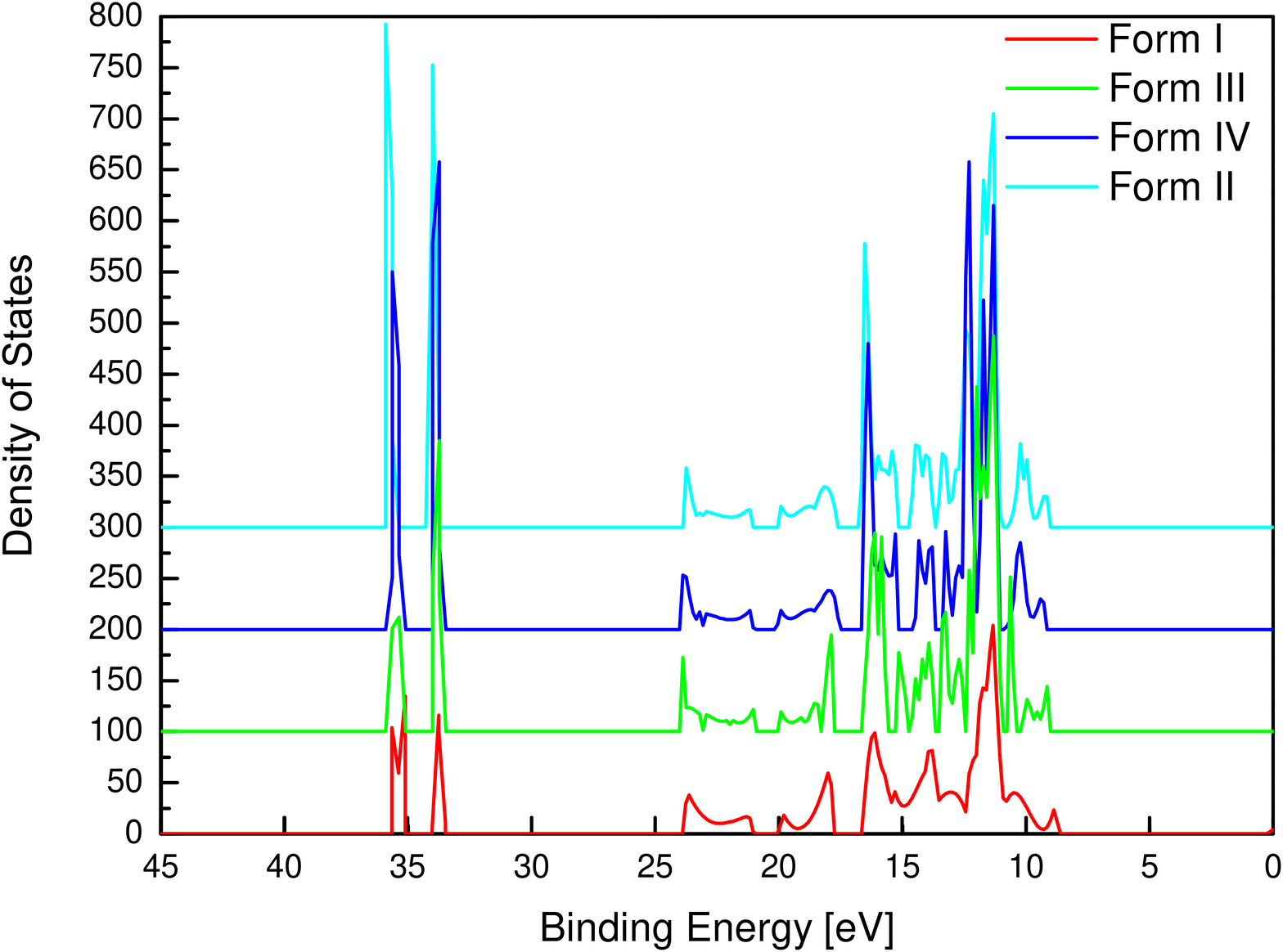}
\end{center}
\caption{Densities of states of the polymorphs of PVDF.}\label{DOS}
\end{figure*}

Density of states of each crystal is shown in Figure \ref{DOS}
To simulate the vXPS of Forms I-IV needs the following: 
the partial DOS curves for each atomic orbital, 
and the theoretical intensity of vXPS used in the Gelius intensity model. \cite{Gel1, Gel2}
The partial DOS curves are extracted from the total DOS.
The theoretical relative photoionization cross-section of each atomic orbital for Al K$\alpha$ radiation determined by Yeh are presented in Table \ref{Yeh}

\begin{table}
\caption{Relative photoionization crss-section of each atomic orbital for H, C and F atoms (relative to C 2s)}\label{Yeh} 
\begin{center}
\begin{tabular}{ccc}
\hline
Atom&Atomic orbital&
\multicolumn{1}{c}{Al K$\alpha$ cross-section} \\
\hline
H&1s&0.0041\\
C&2p&1.0000\\
&2p&0.0323\\
F&2s&4.2797\\
&2p&1.0256\\
\hline
\end{tabular}
\end{center}
\end{table}

\begin{figure*}
\begin{center}
\includegraphics[width=\textwidth]{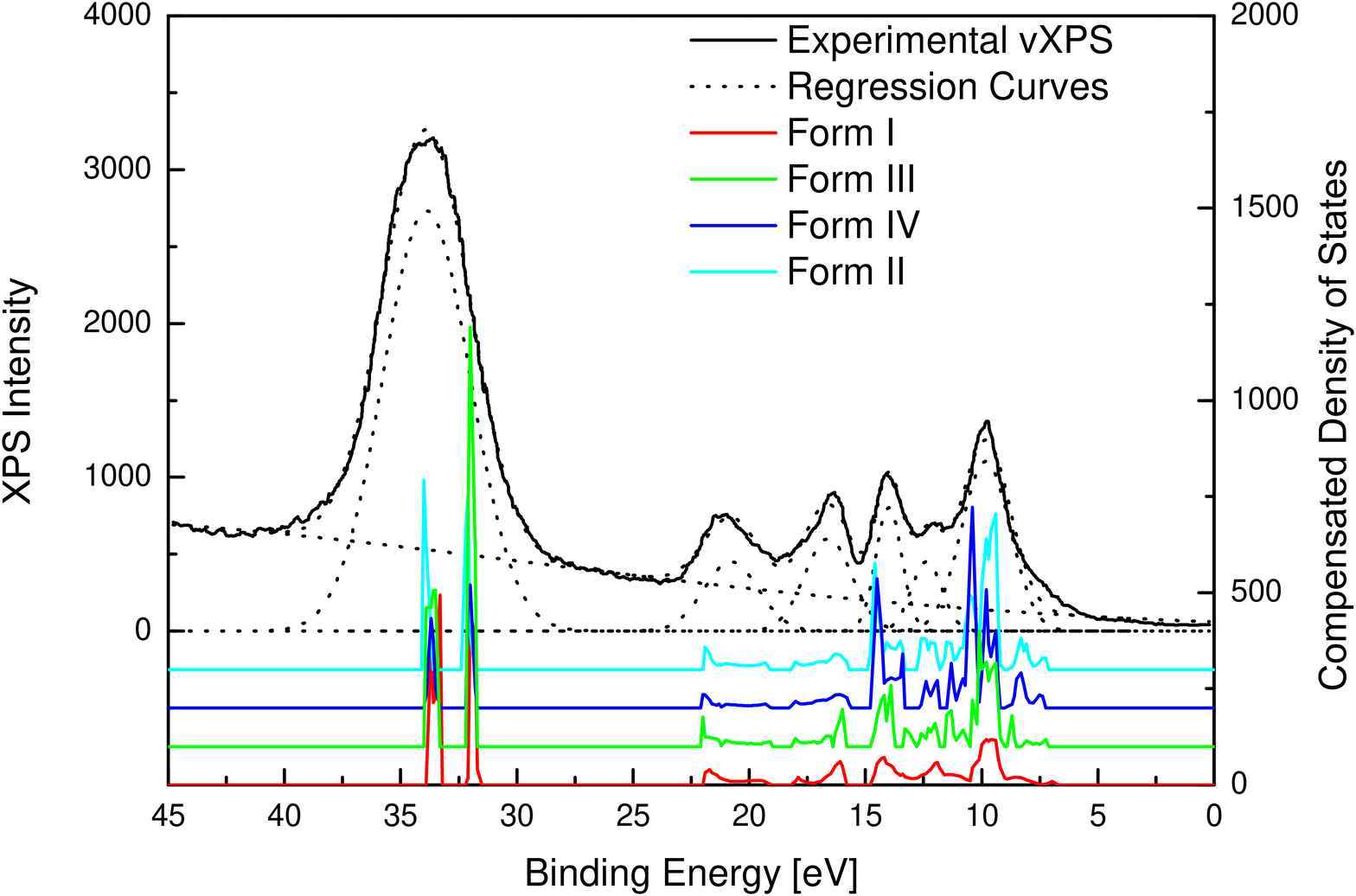}
\end{center}
\caption{A Comparison of experimental vXPS of PVDF and simulated spectra of polymorphs.}\label{vXPS}
\end{figure*}

For Experimental vXPS of PVDF (shown in figure \ref{vXPS}), 
what the intense peaks in the energy range of between 15$\,$eV and 25$\,$ are related to the F2s orbitals of PVDF can be inferred from simulated spectra.
Two broad peaks between $15\,$eV and $25\,$eV are caused by the main C2s and partial F2s and 2p contributions, respectively.
Two intense, broad peaks at lower energy (8--15$\,$eV) are attributed to the F2p contribution to p$\sigma$(C2p-F2p) bonding at around 14$\,$eV, and p lone-pairorbitals at around 10$\,$eV.
The simulated and experimental spectra show good agreement.

\section{Conclusion}
Calculations of all crystal of PVDF were performed employing PBE0/cc-pVTZ.
The lattice constants are calculated by geometry optimization, and are in good agreement.
Difference of the packing energy among the crystals were small.
It can be the reason why the crystals exist in stable.
The author also verified the precision of the molecular structure with the calculation of oligomer, and clarified it.

The possibility of each crystals to become a semiconductor was discussed by the electronic band structure.
The precision of the method employed in the chapter is determined by simulation of vXPS.

The attractive interaction between fluorine atoms in the same CF$_2$ residue is discovered by the analysis of the chemical structure and the band structure.

\onecolumn 

\twocolumn 

\chapter{Physical Properties of Each PVDF Polymorphs}

In this chapter, 
the calculation of harmonic vibrational frequencies at $\Gamma$ point of the polymorphs of PVDF is performed and compared with the experimental infrared/Raman spectrum to examine precision of the PBE0/cc-pVTZ method.
The phonon dispersion of Form I crystal is also calculated, to explain the spectrum.
In addition, the temperature dependency of Gibbs free energy of each crystal is calculated from the frequencies.
Finally, the amounts of spontaneous polarization of each polar crystal are derived employing the well-evaluated hybrid DFT method.

\section{Introduction}

As mentioned in chapter 1, the theoretical prediction of $P_{\rm s}$ of Form I PVDF has been an important goal for many years.
Recently, Nakhmanson \textit{et al.} led the $P_{\rm s}$ to $178\,$mC/m$^2$ using PBE functional with the plane-wave basis set. 
They adopted the experimental lattice constants and mentioned that van der Waals interactions are neglected in DFT.
It is against the philosophy \textit{ab-initio} calculation to rely on experimental value.
And the intermolecular interactions; 
Keesom interaction, the interaction between permanent dipoles;
Debye interaction, the force between permanent dipole and induced dipole;
London dispersion force, instantaneous dipole--induced dipole forces,
can be calculated using some modern quantum chemical method;
moreover the London force is caused by the instantaneous dipole, which is a quantum mechanical phenomenon, and is a consequence of the electronic correlation.

On the other hand, it is known that Forms III and IV of PVDF also have a spontaneous polarization, 
however, the amount is not known,
because the pure and highly oriented sample is not gotten.
The author aimed to confirm the $P_{\rm s}$ of Form I is calculated validly,
and to derive the $P_{\rm s}$ of Form III and IV.

By the way, the normal modes of vibration reflect the strength of the chemical bond in a molecule (in other words, it is resonance of ionic polarization in the theory of dielectrics), are often used to validate the result of calculation
and contribute to Gibbs free energy via excitation energy and entropy.
A chemical potential, the Gibbs energy per mole of each component, is essential quantity to predict whether the phase transition (or the chemical reaction) was able to occur.
According to recent studies,  each crystalline form of PVDF can be produced by controlling the conditions to make a membrane: solvents, temperature and pressure.
Because the mechanism was not yet studied, the explanation by the chemical potential is attempted.

\section{Computational details}
The normal modes of vibrations at $\Gamma$ point for three crystals (Forms I--III) and phonon dispersion for Form I were calculated using the model chemistry of PBE0/cc-pVTZ//PBE0/cc-pVTZ(d,p) and were compared with experimental spectra of precedence study.
Then, the temperature dependency of chemical potential were also estimated using Debye model from the calculated frequencies. 
At last, the amount of polarization for three polar crystals (Forms I, III and IV) were derived with localized Wannier function technique.

The method PBE0/cc-pVTZ//PBE0//cc-pVTZ(d, p) means 
that the structures were derived from the combination of PBE0 hybrid functional and cc-pVTZ basis set (in previous chapter)
and that the modes were calculated employing PBE0 and cc-pVTZ(d, p) which is modified from cc-pVTZ with the polarization functions cc-pVDZ.

A Wannier function $W_n(\vec{r})$ is defined as
\begin{equation}
W_{n \vec R}(\vec r)=\frac{V}{(2\pi)^3}\int\limits_{\rm BZ} \psi_{n \vec k} e^{-i\vec k \cdot \vec R}{\rm d}\vec k,
\end{equation}
where, $\vec{R}$ and $V$ is the position of cell and the volume of unit cells, integration is performed over first Brilloin zone, $\psi_n\vec{k}$ is a Bloch function and remaining variables are defined in chapter 2. Then, the polarization $P$ is described with the elementary charge $e$, $Z_A$ is the atomic number of $A$-th atom and $\vec{R_A}$ of the position of atom as
\begin{equation}
P=\frac eV\sum_A Z_A \vec R_A - \frac{2e}V \sum_n \left<w_n\left|\vec r\right|w_n\right>.
\end{equation}

\section{Results and discussion}
\subsection{Normal modes of vibrations}

The calculated frequencies for Form I PVDF are listed in Table \ref{vibI} with the experimental value. \cite{5vib-I}
And the result for Forms III and II is shown in Table \ref{vibII} and Table \ref{vibIII}, respectively.

\begin{table*}
\caption{Vibrational frequencies at $\Gamma$ point of PVDF Form I. Only$B_2$ mode is IR-inactive, remaining modes are IR and Raman active.}\label{vibI}
\begin{center}
\begin{tabular}{rrcc}
\hline
Calculated &Experimental&Symmetry&Description\\
value [cm$^{-1}$]&value [\rm cm$^{-1}$]\\
\hline
 43&	70&	$B_2$&	Librational mode\\
 260&260& $A_2$&$\tau$ (CF$_2$)\\
 418&	442&	$B_2$&	$\rho$ (CF$_2$)\\
 451&	468&	$B_1$&	$\omega$ (CF$_2$)\\
 510&	508&	$A_1$&	$\delta$ (CF$_2$)\\
 858&	840&	$A_1$&	$\nu_{\rm s}$ (CF$_2$)\\
 897&	880&	$B_2$&	$\rho$ (CH$_2$)\\
 1082&	1071&	$B_1$&	\\
 1189&	1177&	$B_2$&	$\nu_{\rm a}$ (CF$_2$)\\
 1195&\sout{980} 1210&$A_2$&$\tau$ (CH$_2$)\\
 1317&	1273&	$A_1$&	\\
 1428&	1398&	$B_1$&	$\omega$ (CH$_2$)\\
 1454&	1428&	$A_1$&	$\delta$ (CH$_2$)\\
 3133&	2980&	$A_1$&	$\nu_{\rm s}$ (CH$_2$)\\
 3212&	3022&	$B_2$&	$\nu_{\rm a}$ (CH$_2$)\\
 \hline
\end{tabular}\\[0.5pc]
$\tau$: twisting, $\rho$: rocking, $\omega$: wagging, $\delta$: bending, \\$\nu_{\rm s}$: symmetric stretching, $\nu_{\rm a}$: asymmetric stretching
\end{center}
\end{table*}

\begin{table*}
\begin{center}
\caption{Vibrational frequencies of Form III crystal PVDF.}\label{vibIII}
\begin{tabular}{rrrrrr}
\multicolumn{5}{l}{$A'$ mode}\\
\hline
36&
76&
111&
159&
195\\
264&
300&
365&
389&
414\\
477&
545&
449&
786&
828\\
880&
889&
906&
989&
1087\\
1138&
1178&
1192&
1267&
1277\\
1360&
1407&
1434&
1444&
1458\\
3123&
3136&
3182&
3195\\
\hline
\end{tabular}\\[2pc]
\begin{tabular}{rrrrrr}
\multicolumn{5}{l}{$A''$ mode}\\
\hline
17&
44&
105&
133&
222\\
226&
256&
309&
339&
395\\
424&
519&
551&
617&
734\\
802&
854&
888&
902&
1049\\
1067&
1105&
1157&
1223&
1267\\
1307&
1341&
1419&
1438&
1460\\
3124&
3135&
3183&
3194\\
\hline
\end{tabular}
\end{center}
\end{table*}

\begin{table*}
\caption{Vibrational frequencies of Form II crystal PVDF.}\label{vibII} 
\begin{center}
\begin{tabular}{cc}
\multicolumn{2}{c}{IR-active, Raman-Inactive}\\
\begin{tabular}{rrrrr}
\multicolumn{5}{l}{$A_u$ mode} \\
\hline
7&
140&
288&
348&
383\\
536&
773&
869&
887&
1015\\
1189&
1215&
1345&
1412&
1454\\
3136&
3213\\
\hline
\end{tabular}&
\begin{tabular}{rrrrrr}
\multicolumn{5}{l}{$B_u$ mode}\\
\hline
74&
207&
284&
389&
482\\
615&
801&
882&
1110&
1162\\
1266&
1324&
1438&
1461&
3123\\
3220\\
\hline
\end{tabular}\\
\\
\multicolumn{2}{c}{IR-inactive, Raman-active}\\
\begin{tabular}{rrrrr}
\multicolumn{5}{l}{$A_g$} \\
\hline
21&
30&
79&
203&
283\\
384&
486&
618&
812&
886\\
1104&
1162&
1260&
1322&
1441\\
1460&
3116&
3215\\
\hline
\end{tabular}&
\begin{tabular}{rrrrrr}
\multicolumn{5}{l}{$B_g$ mode}\\
\hline
47&
88&
289&
322&
361\\
439&
790&
868&
875&
1004\\
1163&
1244&
1352&
1428&
1457\\
3099&
3202\\
\hline
\end{tabular}\\
\end{tabular}
\end{center}
\end{table*}

\begin{figure*}[]
\includegraphics[width=\textwidth]{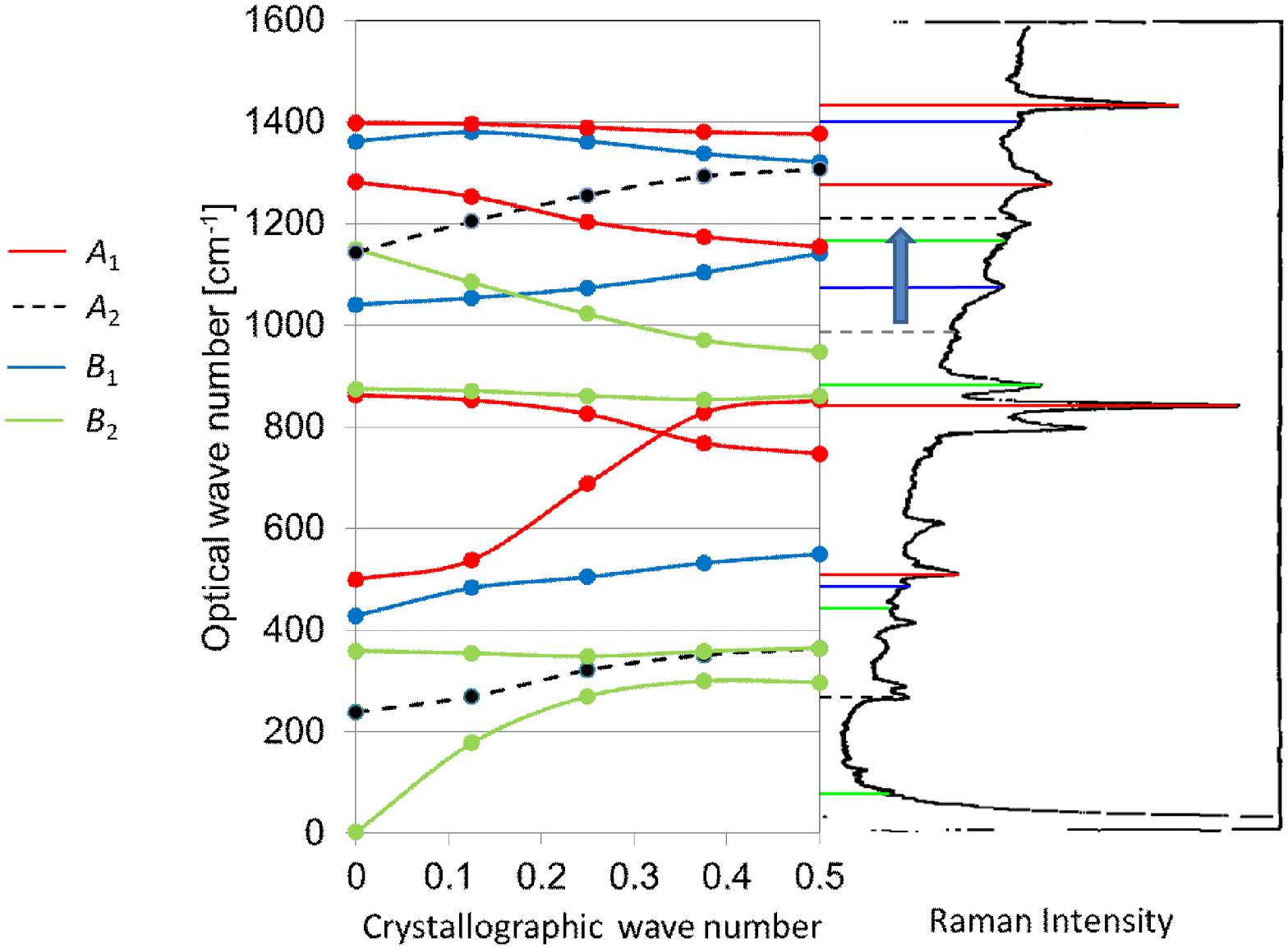}
\caption{Phonon dispersion for Form I PVDF. The left curves are calculated phonon dispersion along $a_3^*$ direction. The right spectrum is Raman spectrum from Reference \cite{5vib-I} with the assignments}\label{phonon-I}
\end{figure*}

For Form I, the calculated vibrational frequency (1195$\,$cm$^{-1}$) of $\tau{\rm (CH_2)}$ mode is shown to be different from the reference value (980$\,$cm$^{-1}$).
To validate this \textit{ab initio} calculation, the phonon dispersion for Form I was calculated with PBE0/cc-pVTZ//PBE0/6-311G** and is shown in figure \ref{phonon-I}.
Van Hove singularities are shown in most phonons at both Gamma point, $k=0$, and Z point, $k=0.5$, 
and corresponded with the peaks in the experimental spectrum.
In other words, the peaks which have no assignments in the article are the van Hove singularities of phonons at Z-point.
Here, a peak at 980$\,$cm$^{-1}$ can be assigned $\nu_{\rm a}{\rm (CF_2)}$ mode which has a $B_2$ symmetry is a green curve shown in the diagram, and
the complicated peaks which can be seen at \textit{ca.} 1210$\,$cm$^{-1}$ can be assigned the complex of $\tau {\rm (CH_2)}$, $A_2$ symmetry, at $\Gamma$ point and $\delta {\rm (CH_2)}$, $A_1$ symmetry, at Z point.
And the handbooks affirm this new assignment.

By the way, in the precedent study, 
it seems that the force fields of molecular mechanic calculation and the assignment were adjusted each together.
It must have been a process:
Initially, the peaks which is easy to assign were chosen in the observed spectrum,
next the force fields were constructed using the frequencies, 
the remaining peaks were assigned with molecular mechanics calculation, 
then the force fields is improved with comparison between the observed and calculated frequencies.
They must have neglected anharmonicity of the $\nu$(C-H) stretching modes and might assumed the modes as the harmonic vibrations, therefore the force field would not be correct.

Except the $\tau{\rm (CH_2)}$ mode and C-H stretching modes, all calculated frequency at $\Gamma$ point with PBE0/cc-pVTZ//PBE0/cc-pVTZ(d,p) matches the precedent assignments.
And the results of Forms III and II are same as the one of Form I;
the same mis-assignments are shown, all frequency except C-H stretching and rocking modes matches.
The scale factor, which is average of experimental value/calcuated value, below 1500$\,$cm$^{-1}$ was 0.9869.

The characteristic bands for Form I are 508$\,$cm$^{-1}$ and 880$\,$cm$^{-1}$ in the experiment, 
and calculated values are 510$\,$cm$^{-1}$ and 897$\,$cm$^{-1}$, respectively.
For Form III, the experimental value is 442$\,$cm$^{-1}$ and calculated value is 414$\,$cm$^{-1}$. \cite{5vib-III}
For Form II, 531$\,$cm$^{-1}$, 612$\,$cm$^{-1}$,766$\,$cm$^{-1}$,795$\,$cm$^{-1}$ and 976$\,$cm$^{-1}$ in the reference become 536$\,$cm$^{-1}$, 615$\,$cm$^{-1}$, 773$\,$cm$^{-1}$,801$\,$cm$^{-1}$ and 1015$\,$cm$^{-1}$ in the calculation, respectively. \cite{5vib-3}
These are in good agreement enough.

\subsection{Temperature dependencies of chemical potential} 
\begin{figure*}
\includegraphics[width=\textwidth]{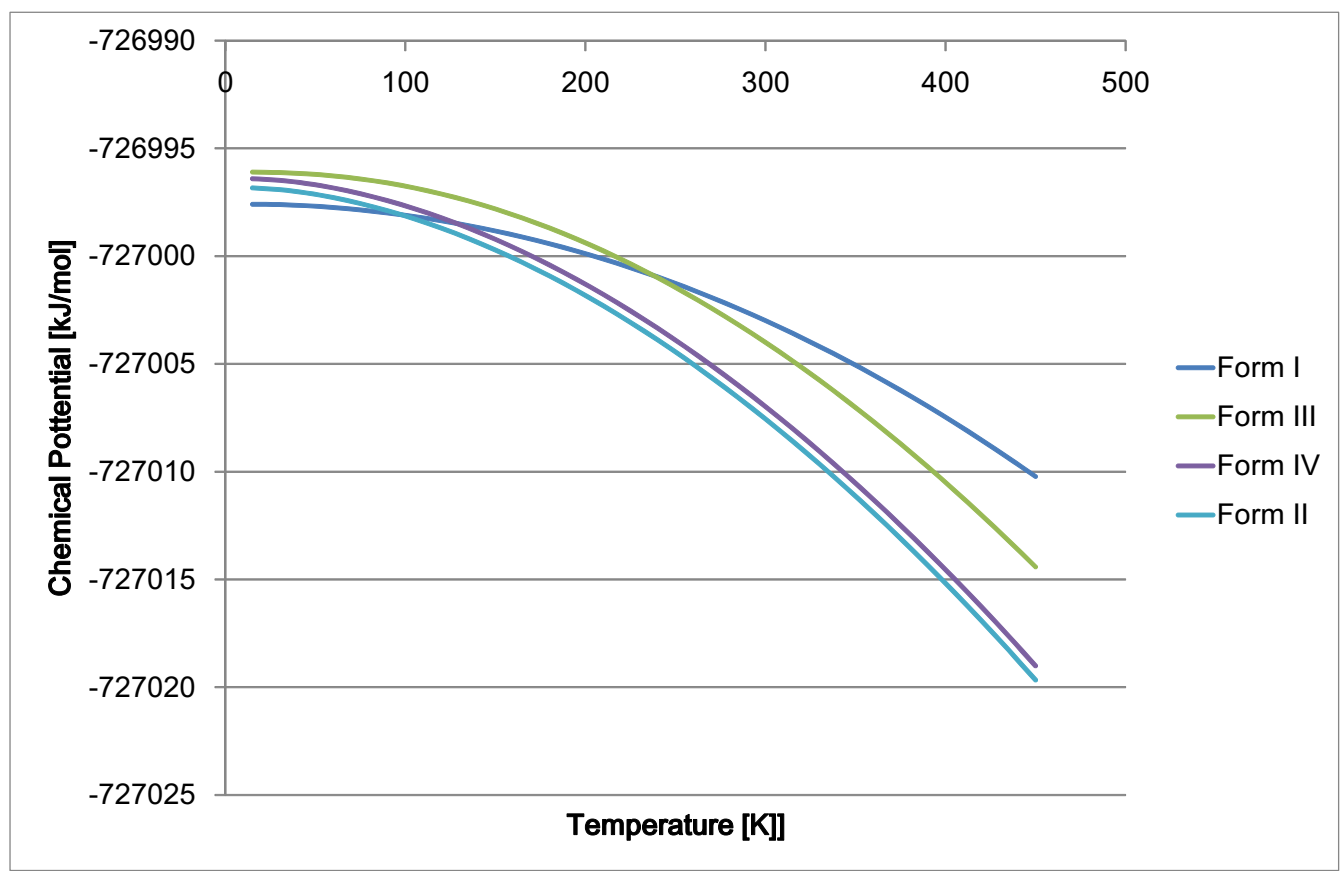}
\caption{Temperature dependencies of chemical potentials of PVDF polymorphs.}\label{cp}
\end{figure*}

The temperature dependency from 15$\,$K to 450$\,$K of chemical potential of each crystal of PVDF is plotted in figure \ref{cp}
The calculations were performed employing PBE0/cc-pVTZ//PBE0/cc-pVTZ(d,p) with Debye model.

At the higher temperature, the chemical potentials indicate the lower energies.
The chemical potential of Form II is the lowest at high temperature below melting point, 177\,$\rm{^\circ\!C}$ among the four crystals.
At the low temperature, 
the potential of Form I is the lowest,
it is suspected that entropy and excitation energy those derived from the librational and lattice modes, 
which correlate with the rotational modes of isolated molecule, 
of each crystal are involved with the result.
The cross-point is qualitative because of the following reasons:
The estimation is very rough, the curves are steep, and the acoustic modes are neglected.
However, Form II is produced with melt-crystallization process.
And it seems that some recent studies has shown that Form I crystal tends to be produced at relatively low temperature (room temperature or lower). \cite{5Horibe}

\subsection{Spontaneous polarization}
The amount of spontaneous polarization $P_{\rm s}$ of each crystal perpendicular to the chain axis  was evaluated through Wannier functions.
For Form I, $P_{\rm s}$ was derived to 176$\,$mC/m$^2$.
$P_{\rm s}$ for Form III and IV were 71$\,$mC/m$^2$ and 85$\,$mC/m$^2$, respectively
To analyze in detail, calculations about atomic charge were performed (Table \ref{charge-I}).
The charge of quantum mechanical calculations was derived from Mulliken population analysis. 
The charge for the oligomer were calculated with MP2/aug-cc-pVTZ,
the calculations for infinite chain and crystal were performed with PBE0/cc-pVTZ.

\begin{table}
\caption{Atomic charge of Form I PVDF through Mulliken population analysis.}\label{charge-I}
\begin{center}
\begin{tabular}{p{0.05\textwidth}p{0.1\textwidth}p{0.1\textwidth}p{0.1\textwidth}}
\hline
&
\multicolumn{1}{c}{MP2}&
\multicolumn{2}{c}{PBE0}\\
\cline{2-4}
&
\multicolumn{1}{c}{Oligomer}&
\multicolumn{1}{c}{Chain}&
\multicolumn{1}{c}{Crystal}\\
\hline
$\rm C_H$&
\multicolumn{1}{r}{$-0.284$}&
\multicolumn{1}{r}{$-0.254$}&
\multicolumn{1}{r}{$-0.481$}\\
H&
\multicolumn{1}{r}{$0.121$}&
\multicolumn{1}{r}{$0.105$}&
\multicolumn{1}{r}{$0.289$}\\
$\rm C_F$&
\multicolumn{1}{r}{$0.491$}&
\multicolumn{1}{r}{$0.381$}&
\multicolumn{1}{r}{$0.273$}\\
F&
\multicolumn{1}{r}{$-0.191$}&
\multicolumn{1}{r}{$-0.172$}&
\multicolumn{1}{r}{$-0.185$}\\
\hline
&&&\\
\end{tabular}
\end{center}
\end{table}

The result obtained for Form I is almost the same as which reported by Nakhmanson \textit{et al.} \cite{5Nakhmanson04, 5Nakhmanson05}
even though their calculation method was different.
As mentioned in Chapter 1, rigid dipole model gives $P_{\rm s}$ of \textit{ca.} 130$\,$mC/m$^2$ for Form I.
The larger value of the calculated $P_{\rm s}$ suggests that the polarization is enhanced by 35\% in the Form I crystal; 
that is, the spontaneous polarization is stabilized by Coulomb interactions, resulting in ferroelectric cooperative effect. 

Total sum of the atomic charge in oligomer (2,2,4,4,6,6-hexafluoroheptane) is not zero,
because only charge in the center residues (CH$_2$ of 3rd locant and CF$_2$ of 4th locant) is shown. 
The atomic charge shows the enhancement of polarity of the C-H bonding in the crystal.
The deficiency of electron at a hydrogen atom is calculated to be 0.289, that is about twice of the amounts of oligomer and chain (0.121 and 0.105).
The charge of carbon atoms seem to be re-distributed, 
the charge of fluorine atoms about $-0.2e$ is almost same among the results.
Thus, the enhancement of $P_{\rm s}$ in Form I crystal can be attributed, at least partially, to the enhanced polarity of the C-H bondings. 
In other words, 
the local field induced in crystal is parallel to the dipole direction, 
therefore the $P_{\rm s}$ is enhanced from classical prediction; 
and the dipole moment is increased from isolated one.
On the other hand, the field in chain is against the dipoles; the polarization is decreased.
Simple estimation of local field is done by counting the dipoles;
the horizontal dipoles, which can be seen in chain direction and the direction perpendicular to the chain and dipole direction, must be weaken each other,
and the vertical dipoles, along the polarization direction, must be increased each together.
In addition, similar enhancement is found to occur in crystalline Forms III and IV with Mulliken population analysis.

Because samples consisting of only Form III or IV crystal cannot be obtained for PVDF,
$P_{\rm s}$ for these crystalline forms has not been determined experimentally, 
and no prediction has been seen. 
Thus, this is the first report estimating the values of $P_{\rm s}$ for Forms III and IV of PVDF.
$P_{\rm s} = 85\,{\rm mC/m^2}$ of Form IV is nearly half that of Form I and is consistent with the crystal structure because the chain includes gauche conformation and the dipoles incline by 60$^\circ$ from the direction of total polarization. 
In contrast, $P_{\rm s}$ of Form III is much smaller even though the inclination of dipoles is the same as that of Form IV.
This may be related to the difference of ferroelectric cooperative effect in these two polymorphs.

\section{Conclusions} 
The calculation of the normal modes of vibrations at $\Gamma$ point of PVDF polymorphs of Forms I, III and II was performed employing PBE0/cc-pVTZ//PBE0/cc-pVTZ(d,p) and was compared with experimental spectra, 
the result is that the characteristic bands are in agreement well.
The calculation of phonon dispersion of Form I was performed, and reproduces the Raman spectrum.
With the vibrational frequencies, the temperature dependency of chemical potential of each crystal is evaluated through Debye model.
Considering the precision of PBE0/cc-pVTZ, spontaneous polarization for each polar crystal is derived.

The results of vibrational frequencies at $\Gamma$ point and phonon dispersion were disagreed with the past study partially and invite to new assignment of IR/Raman spectra.
The new assignments for Form I PVDF were established through detailed analysis of phonon dispersion and had a different $\tau {\rm CH_2}$ value from past study, new one was observed at 1210$\,$cm$^{-1}$. 
Assignments of frequencies of the remaining crystals will be done.

The temperature dependency of chemical potential was that Form II crystal had the lowest chemical potential at high temperature and gives an explanation that Form II crystal is produced with melt-crystallization process.
The result at the low temperature meant that Form I crystal had the lowest free energy, 
so Form I crystal may have an advantage to be produced at relatively low temperature.

Considering the verifications (crystal structures, electronic state and vibrational frequencies) until now, the spontaneous polarization of each polar crystal, Forms I, III and IV was derived to $176,{\rm mC/m^2}$, $71,{\rm mC/m^2}$ and $85,{\rm mC/m^2}$ through the Wannier functions with the method PBE0/cc-pVTZ which is examined comprehensively for PVDF and is worth precise enough.
The spontaneous polarization behavior of these crystals strongly demonstrates the importance of precise quantum chemical treatment to examine electric properties, including ferroelectricity.

These results indicate that the improvement of film deposition process will enhance the physical properties derived from ferroelectricity of PVDF.

\onecolumn 

\twocolumn 
 
\chapter{Conclusions}

In this thesis, 
the aim is to derive spontaneous polarization of each polymorph of PVDF,
the method is DFT method which is established through the validation of the crystallographic, molecular and electronic structure of Form I PVDF and is used in CRYSTAL09 under three dimensional periodic boundary conditions,
the results are followings:
\begin{enumerate}
\item The most suitable method was PBE0/cc-pVTZ to analyze theoretically the structure and the optical properties of the crystals of PVDF.
\item The origin of polymorphism of PVDF was determined.
\item The amount of spontaneous polarization of all polar crystal of PVDF were derived from \textit{ab initio} calculation.
\end{enumerate}

This thesis includes six chapters:

In chapter 1, the introduction of PVDF and the purpose of this study were described.
PVDF is a ferroelectric polymer and has crystal polymorphs.
The theoretical prediction of its spontaneous polarization of Form I PVDF has been the important goal for long years.
However, the remaining polar crystals were neglected.

In chapter 2, the quantum chemical methods, mainly about DFT method, which were employed through this thesis were described.

In chapter 3, 
to determine the characteristics of Form I PVDF, 
various methods were exhaustively evaluated by the crystal,  molecular and electronic structure, 
and the suitable method, PBE0/cc-pVTZ means the combination of PBE0 functional and cc-pVTZ basis set, shows the best agreement of lattice constants with experimental ones and shows the reasonable total energy.

In chapter 4,
the method, PBE0/cc-pVTZ, was adopted for four polymorphs of PVDF.
The packing energy, the electronic band structures and density of states were derived.
With the electronic state, valence X-ray spectra are simulated.
In the result, the calculated lattice constants of each crystal are well in agreement with experimental values;
the deviation was lower than 1\% in the chain direction of $c$ axis,
the deviations were lower than 5\% in the intermolecular direction $a$ and $b$ axes.
The difference of energy derived from the total packing energy was less than $3\,$kJ/mol and is regarded as the origin of crystalline polymorphism.
The molecular structure and the electronic band structures derive the evidence of the weak attractive interaction between Fs in CF$_2$.
The band structure of Form I suggests the $n$-type semiconductivity.
Finally, the simulated valence X-ray spectra were derived from the electronic structures and are matched with the experimental spectrum.

In chapter 5,
the normal modes of vibrations for Forms I--III were calculated employing PBE0/cc-pVTZ//PBE0/cc-pVTZ(d,p) and agree with the experimental IR/Raman spectra,
based on the inspection above, 
the calculations about spontaneous polarizations of the polar crystals, Forms I, III and IV, were performed at PBE0/cc-pVTZ level;
the amount of Form I was greater than the one of rigid dipole model,
therefore, it seems that the ferroelectric interaction, which derived from the Coulomb interaction and stabilize the electric dipoles each other, contributed to the mechanism to build the spontaneous polarization of PVDF crystals,

As mentioned above, in this study, 
calculations for each polymorph of PVDF were performed exhaustively and accurately,
the molecular structure in crystals and the detail of chemical bonds are clarified,
a new assignment of the vibrational spectra was gotten as a result.
For the first time, the amount of spontaneous polarization of all crystals are derived.

The computational method not only contributes to evaluation of ferroelectric properties of PVDF but also is able to be applied in the broad field of the solid-state science and the condensed matter.
In another view to the application,
The enhancement of the process, which must be done by controlling the high-order structures, to deposit films will lead to make a sample which have greater ferroelectric characterizations.

As the further prospective, the static comprehensions of the crystals were finished, the molecular dynamics will be applied to the PVDF and will study the mechanism of crystallization and the behavior of polarization reversal.

\onecolumn
\chapter*{Acknowledgments}
I would like to express my special thanks of gratitude to my teachers:
Prof. Hirofumi Yajima, who was the supervisor at doctoral course, tutored me about chemistry of various materials and optical theories of materials.
Prof. Takeo Furukawa, who was the advising teacher at master's course, instructed me in the theories of poly(vinylidene fluoride) and the essence of electronics also after he was transfered to Kobayashi Institute of Physical Research.
Prof. Kazunaka Endo, who was a visiting professor in Tokyo University of Science, taught me about XPS and quantum theories.
Kohji Tsuchiya, who was assistant teacher at the Yajima laboratory, kept the comfortable research environment.
Yoshiyuki Takahashi was assistant teacher at the Furukawa Laboratory and helped me to submit article to Polymer Journal.

In addition, the high-performance computing resouce was provided by Tokyo University of Science.

\end{document}